\documentclass[12pt,draftcls,onecolumn]{IEEEtran}

\usepackage{etex}
\usepackage{bbm}
\usepackage{balance}
\usepackage{cite}
\usepackage{soul}
\usepackage{epstopdf}
\usepackage{amssymb}
\usepackage[cmex10]{amsmath}
\usepackage{cases}
\let\labelindent\relax
\usepackage{enumitem}
\usepackage{array}
\usepackage{multirow}
\usepackage{bigdelim}
\usepackage{mdwtab}
\usepackage{tikz}
\usepackage{physics}
\usepackage{color}
\usepackage{booktabs}

\makeatletter
\let\MYcaption\@makecaption
\makeatother
\usepackage[font=footnotesize]{subcaption}
\makeatletter
\let\@makecaption\MYcaption
\makeatother

\DeclareMathOperator*{\arginf}{arg\,inf}

\usepackage{accents}

\newtheorem{theorem}{Theorem}[section]
\newtheorem{lemma}{Lemma}[section]
\newtheorem{proposition}{Proposition}[section]
\newtheorem{corollary}{Corollary}[section]

\newtheorem{remark}{Remark}[section]

\title{Smoother Entropy for Active State Trajectory Estimation and Obfuscation in POMDPs}%
\author{Timothy L.\ Molloy, \IEEEmembership{Member, IEEE}, and Girish N.\ Nair, \IEEEmembership{Fellow, IEEE}%
\thanks{The first author was with the Dept.\ of Electrical and Electronic Engineering, University of Melbourne, VIC 3010, Australia. He is now with the CIICADA Lab, School of Engineering, Australian National University (ANU), Canberra, ACT 0200, Australia (e-mail: timothy.molloy@anu.edu.au) The second author is with the Dept.\ of Electrical and Electronic Engineering, University of Melbourne, VIC 3010, Australia (e-mail: gnair@unimelb.edu.au)}
\thanks{This work received funding from the Australian Government, via grant AUSMURIB000001 associated with ONR MURI grant N00014-19-1-2571.}%%
\thanks{Preliminary versions of some results in this paper were presented at the 2021 American Control Conference \cite{Molloy2021} and the 2021 European Control Conference \cite{Molloy2021a}.}%
}

\begin{document}

\maketitle
\thispagestyle{empty}

\begin{abstract}
\boldmath
We study the problem of controlling a partially observed Markov decision process (POMDP) to either aid or hinder the estimation of its state trajectory.
We encode the estimation objectives via the \emph{smoother entropy}, which is the conditional entropy of the state trajectory given measurements and controls.
Consideration of the smoother entropy contrasts with previous approaches that instead resort to marginal (or instantaneous) state entropies due to tractability concerns.
By establishing novel expressions for the smoother entropy in terms of the POMDP belief state, we show that {both the problems of minimising and maximising the smoother entropy in POMDPs can surprisingly be reformulated as belief-state Markov decision processes with concave cost and value functions.
The significance of these reformulations is that they render the smoother entropy a tractable optimisation objective, with structural properties amenable to the use of standard POMDP solution techniques for \emph{both} active estimation and obfuscation.}
Simulations illustrate that optimisation of the smoother entropy leads to superior trajectory estimation and obfuscation compared to alternative approaches.
\end{abstract}

\begin{IEEEkeywords}
Partially observed Markov decision process (POMDP), entropy, estimation, directed information.
\end{IEEEkeywords}

% \fi
\IEEEpeerreviewmaketitle

\section{Introduction}
The problem of controlling a stochastic dynamical system to either aid or hinder the estimation of its time-varying state arises across numerous applications in automatic control, signal processing, and robotics.
Applications in which the problem has been investigated in its \emph{active estimation} form to aid state estimation include active state estimation and dual control in automatic control \cite{Blackmore2008, Hu2004,Baglietto2007,Scardovi2007}, controlled sensing in signal processing and robotics \cite{Krishnamurthy2016,Zois2014,Zois2017,Chattopadhyay2018,Krishnamurthy2020,Hoffmann2010,Kartik2018}, and active simultaneous localisation and mapping (SLAM) in robotics \cite{Mu2016, Thrun2005, Roy1999a, Stachniss2005, Valencia2012,Roy2005}.
Conversely, applications in which the problem has been investigated in its \emph{active obfuscation} form to hinder state estimation include privacy in cyber-physical systems \cite{Li2018,Li2019,Farokhi2020,Tanaka2017,Savas2020,Shateri2020}, and covert navigation in robotics \cite{Marzouqi2011,Hibbard2019}.
Despite these many applications, few works have explicitly addressed active estimation or obfuscation of entire \emph{state trajectories}, with most instead focusing on aiding or hindering state estimation as it relates to the performance of Bayesian filters.
Bayesian filters provide marginal state estimates given a history of observations and controls.
However, in many applications such as target tracking and SLAM, (joint) state trajectory estimates are of greater interest than marginal state estimates.
For instance, in surveillance applications, it can be important to estimate or conceal not just where a target currently is, but from where it came and what points it visited.
Similarly in SLAM, better estimates of the past robot trajectory help reconstruct a more accurate map of the environment.
Motivated by such applications, in this paper we investigate novel approaches to active state estimation and obfuscation that explicitly relate to estimating or concealing entire state trajectories.

\subsection{Related Work}

Developing meaningful measures of state uncertainty (or estimation performance) that are tractable to optimise within standard stochastic optimal control frameworks such as partially observed Markov decision processes (POMDPs) is a key challenge in active estimation and obfuscation.
The solution of standard POMDPs involves reformulating them as fully observed Markov decision processes (MDPs) in terms of a belief (or information) state corresponding to the state estimate provided by a Bayesian filter.
Numerous algorithms exist for solving the resulting belief-state MDPs, with the vast majority relying on the fact that standard POMDPs have cost and value (or cost-to-go) functions that are concave or piecewise-linear concave (PWLC) in terms of the belief state (see \cite{Haugh2020,Walraven2019,Krishnamurthy2016,Kurniawati2008,Garg2019,Krishnamurthy2007} and references therein).
The intrinsic relationship between Bayesian filters and belief-state approaches for solving POMDPs has resulted in state-uncertainty measures related to filter estimates dominating the literature of both active state estimation and obfuscation (see \cite{Krishnamurthy2007,Krishnamurthy2016,Krishnamurthy2020,Araya2010,Li2019} and references therein) --- with particular interest paid to state-uncertainty measures that are concave or PWLC functions of the belief state (cf.\ \cite{Araya2010} and \cite[Chapter 8]{Krishnamurthy2016}).

State-uncertainty measures previously considered for active estimation include the error probabilities \cite{Krishnamurthy2007,Blackmore2008}, mean-squared error \cite{Zois2014,Zois2017,Krishnamurthy2007}, Fisher information \cite{Flayac2017}, {expected confidence \cite{Kartik2018}}, and entropy \cite{Krishnamurthy2007,Scardovi2007,Roy1999a,Thrun2005} of Bayesian filter estimates (see also \cite[Chapter 8]{Krishnamurthy2016} and references therein).
Similarly, active obfuscation approaches such as \cite{Li2019} consider minimising the probability mass of filter estimates at the true states.
Unfortunately, these popular state-uncertainty measures based on filter estimates are of limited use in describing and optimising the uncertainty associated with entire {(time-varying)} state trajectories, since they neglect temporal correlations between states that arise due to the state dynamics.
Without consideration of temporal correlations, active estimation approaches may select actions that lead to highly random (or uncertain) state transitions, and active obfuscation approaches such as \cite{Li2019} leave open the possibility of adversaries accurately inferring states at isolated times and using correlations to estimate the entire trajectory via Bayesian smoother-like algorithms (e.g., fixed-interval Bayesian smoothers and the Viterbi algorithm, cf.\ \cite[Section 3.5]{Krishnamurthy2016}).

Bayesian smoother-like algorithms are concerned with inferring the states of partially observed stochastic systems given entire measurement and control trajectories.
Unlike Bayesian filters, they are capable of exploiting correlations between past, present, and future measurements and controls to compute state estimates (cf.\ \cite[Section 3.5]{Krishnamurthy2016}).
Bayesian smoother-like algorithms have been studied over many decades and constitute key components in many target tracking (cf.~\cite{Bar-Shalom2001}) and robot SLAM (cf.~\cite{Thrun2005}) systems.
The problem of controlling a system so as to either aid or hinder the estimation of its state trajectory with smoother-like algorithms has received limited attention, with most efforts confined to the robotics literature on active SLAM (cf.~\cite{Mu2016,Stachniss2005,Valencia2012}).
Treatments in robotics have, however, avoided the use of state-uncertainty measures related to trajectories due to tractability concerns, and have instead resorted to sums of marginal (or instantaneous) state-uncertainty measures without consideration of temporal state correlations between states (cf.\ \cite{Stachniss2005,Valencia2012}).
Indeed, few state-uncertainty measures explicitly related to entire trajectories or trajectory estimates have been investigated for active estimation.

Most recently, the problem of obfuscating entire state trajectories from \emph{any} conceivable estimator has been investigated by drawing on ideas from privacy in static settings (e.g., datasets) including differential privacy \cite{Sandberg2015,Farokhi2020,Hale2015} and information theory \cite{Nekouei2019,Farokhi2020,Tanaka2017,Murguia2021}.
%For example, the mutual information \cite{Murguia2021}, directed information \cite{Tanaka2017,Nekouei2019,Shateri2020}, and Fisher information \cite{Farokhi2018} between state and measurement trajectories have all been considered.
These works, however, sidestep complete POMDP treatments either by only increasing the state's unpredictability \cite{Savas2020, Hibbard2019} or by only degrading the measurements \cite{Nekouei2019,Tanaka2017,Murguia2021} (rather than a combination of the two).
Furthermore, as noted in \cite{Fehr2018}, POMDPs for information-averse or obfuscation problems frequently involve cost and value functions that are not concave in the belief state, and so may have been mostly avoided until recently because no satisfying (approximate) solution techniques existed.

\subsection{Contributions}
In this paper, we investigate the conditional entropy of the state trajectory given measurements and controls as a \emph{tractable} state-uncertainty measure for both active state estimation and obfuscation in POMDPs.
We dub this conditional entropy the \emph{smoother entropy} since it plays a pivotal role in tight upper and lower bounds on the minimum achievable probability of error for any conceivable state-trajectory estimator (cf.\ \cite{Feder1994}), including Bayesian smoother-like algorithms.
Prior literature has dismissed the smoother entropy as an intractable objective in POMDPs (cf.\ \cite{Stachniss2005,Valencia2012}), since it has not been shown to be a function of the POMDP belief state with structural properties (e.g.\ additivity and concavity in the belief state) amenable to the use of standard POMDP solution techniques (e.g., dynamic programming).
However, by using the Marko-Massey theory of \emph{directed information} \cite{Marko1973,Massey1990,Kramer1998,Massey2005}, we show that there are multiple belief-state forms of the smoother entropy, with one form leading to a belief-state MDP reformulation of active state estimation with concave cost and value functions, and another leading to a belief-state MDP reformulation of active state obfuscation with concave cost and value functions.
These concavity results are surprising since active estimation involves minimising the smoother entropy whilst active obfuscation involves maximising it, and POMDP formulations of obfuscation have frequently been avoided due to non-concave cost and value functions (cf.\ \cite{Fehr2018}).
They are also practically important since they enable the use of standard POMDP solution techniques.

The key contributions of this paper are:
\begin{enumerate}
    \item
    The derivation of two novel expressions for the smoother entropy in POMDPs in terms of the POMDP belief state, through the use of the Marko-Massey theory of directed information; and,
    \item
    The surprising demonstration that both the problems of minimising and maximising the smoother entropy in POMDPs can be formulated as belief-state MDPs with concave cost and value functions, using our novel expressions for the smoother entropy.
\end{enumerate}
The practical significance of these contributions is that they render the smoother entropy a tractable objective in POMDPs for both active state estimation and active state obfuscation with structural properties amenable to the use of standard POMDP solution techniques.
We specifically present a bounded-error dynamic programming solution technique based on PWLC approximations of the cost and value functions for either minimising the smoother entropy (for active state estimation) or maximising it (for active state obfuscation).

Compared to our early work in \cite{Molloy2021,Molloy2021a}, significant extensions in this paper include: 1) Use of the Marko-Massey theory of directed information to unify the derivations of belief-state smoother entropy forms and enable comparison with the directed-information work of \cite{Tanaka2017,Nekouei2019}; 2) Characterisation of the structural properties of all belief-state MDP formulations of our active estimation and obfuscation problems; 3) Development of PWLC (approximate) solutions and their associated error bounds; and 4) Numerical and theoretical analysis examining the operational relationship between smoother-entropy optimisation and estimation error probabilities.
%With the exception of Lemma \ref{lemma:estConcave} and Theorem \ref{theorem:estConcave} (published in \cite{Molloy2021a} without detailed proofs), the technical results of this paper are new in their full generality.

\subsection{Paper Organisation}
This paper is structured as follows.
In Section \ref{sec:problem}, we introduce the smoother entropy and an active estimation or obfuscation problem involving its minimisation or maximisation.
In Section \ref{sec:directedInformation}, we establish novel additive and belief-state forms of the smoother entropy.
In Section \ref{sec:activeEst}, we exploit our smoother entropy forms to reformulate our active estimation or obfuscation problem as belief-state MDPs, examine the structure of these belief-state MDPs, and use their structure to develop an approach to finding bounded-error solutions to them via standard POMDP techniques.
In Section \ref{sec:operational}, we discuss the operational significance of our smoother entropy results.
Finally, we illustrate our results in simulations in Section \ref{sec:results} and provide conclusions in Section \ref{sec:conclusion}.

%inspired by privacy in cloud-based control (e.g., \cite{Tanaka2017}) and uncertainty-aware robot navigation (e.g., \cite{Roy1999, Thrun2005,Nardi2019}) in Section \ref{sec:results}. We provide conclusions in Section \ref{sec:conclusion}.

\subsection{Notation}
Random variables will be denoted by capital letters, and their realisations by lower case letters (e.g., $X$ and $x$).
Sequences of random variables and their realisations will be denoted by capital and lower case letters, respectively, with superscripts denoting their final index (e.g., $X^T \triangleq \{X_0, X_1, \ldots, X_T\}$ and $x^T \triangleq \{x_0, x_1, \ldots, x_T\}$).
With a mild abuse of notation, the probability mass function (pmf) of a random variable $X$ (or its probability density function if it is continuous) will be written as $p(x)$, the joint pmf of $X$ and $Y$ as $p(x, y)$, and the conditional pmf of $X$ given $Y = y$ as $p(x|y)$ or $p(x | Y = y)$.
For a function $f$ of $X$, the expectation of $f$ evaluated with $p(x)$ will be denoted $E_X [f(x)]$ (i.e., random variables in expectations will be denoted by lower case letters). 
The conditional expectation of $f$ evaluated with $p(x|y)$ will be similarly denoted $E[f(x) | y]$.
With a common abuse of notation, $E_\mu [ \cdot ]$ is also used to indicate the dependence of an expectation on a policy $\mu$.
The {\em pointwise} (discrete) entropy of $X$ given $Y = y$ will be written $H(X | y) \triangleq - \sum_{x} p(x|y) \log p(x|y)$ with the (average) conditional entropy of $X$ given $Y$ being $H(X | Y) \triangleq E_{Y} \left[ H(X|y) \right]$.
The mutual information between $X$ and $Y$ is $I(X; Y) \triangleq H(X) - H(X | Y) = H(Y) - H(Y | X)$.\footnote{If $Y$ is continuous-valued, then $H(Y)$ ($H(Y|X)$) is replaced with the \emph{differential entropy} $h(Y)$ (resp.\ \emph{conditional differential entropy} $h(X|Y)$) \cite{Cover2006}.}
The pointwise conditional mutual information of $X$ and $Y$ given $Z = z$ is $I(X; Y | z) \triangleq H(X | z) - H(X | Y, z)$ with the (average) conditional mutual information given by $I(X; Y | Z) \triangleq E_{Z} \left[ I(X; Y | z) \right]$.
Where there is no risk of confusion, we will omit the adjectives ``pointwise'' and ``conditional''.
Where entropies and mutual informations are associated with expectations involving a policy $\mu$, a subscript will be used (e.g., $H_\mu(X)$ and $I_\mu(X;Y)$). 

\section{Problem Formulation and Solution Approach}
\label{sec:problem}

In this section, we formulate an active state estimation or obfuscation problem using the smoother entropy, and sketch our approach to solving it as a POMDP.

\subsection{Active Estimation or Obfuscation Problem Formulation}
Let $X_k$ for $k \geq 0$ be a discrete-time first-order Markov chain with a finite state space $\mathcal{X} \triangleq \{1, 2, \ldots, N_x\}$.
Let the initial pmf of $X_0$ be the vector $\rho \in \Delta$ with components $\rho(x) \triangleq P(X_0 = x)$ for $x \in \mathcal{X}$.
The initial pmf belongs to the $(N_x-1)$-dimensional probability simplex $\Delta \triangleq \{\pi \in [0,1]^{N_x} : \sum_{x \in \mathcal{X}} \pi(x) = 1\}$.
We shall let the (controlled) transition dynamics of $X_k$ be described by:
\begin{align}
    \label{eq:stateProcess}
    A^{x,\bar{x}}(u) \triangleq p( X_{k+1} = x | X_k = \bar{x}, U_k = u)
\end{align}
for $k \geq 0$ with the controls $U_k$ belonging to the finite set $\mathcal{U} \triangleq \{1, 2, \ldots, N_u\}$.
The state process $X_k$ is (partially) observed through a stochastic measurement process $Y_k$ for $k \geq 0$ taking values in a (potentially continuous) metric space $\mathcal{Y}$.
The measurements $Y_k$ are distributed according to:
\begin{align}
    \label{eq:obsProcess}
    B^{x,y} (u) \triangleq p( Y_k = y | X_k = x, U_{k-1} = u)
\end{align}
for $k > 0$ with $B^{x_0,y_0} \triangleq p( Y_0 = y_0 | X_0 = x_0 )$ and where the kernel \eqref{eq:obsProcess} is a conditional probability density function (pdf) when $\mathcal{Y}$ is continuous, and a conditional pmf when $\mathcal{Y}$ is finite.

The controls $U_k$ for $k \geq 0$ are given by a potentially stochastic output-feedback policy $\mu \triangleq \{ \mu_k^{i_k} : k \geq 0\}$ described by (conditional) pmfs
\begin{align*}
    \mu_k^{i_k}(u_k) 
    \triangleq p(U_k = u_k | Y^k = y^k, U^{k-1} = u^{k-1})
\end{align*}
where $i_k \triangleq (y^{k},u^{k-1})$ is a realization of the \emph{information state} $I_k \triangleq (Y^k, U^{k-1})$.
A policy $\mu = \{ \mu_k^{i_k} : k \geq 0\}$ will be said to be \emph{deterministic} if, at all times $k \geq 0$, the support of $\mu_k^{i_k}$ is concentrated at a single control $u_k$; otherwise $\mu$ is \emph{stochastic}.
We shall denote the set of all policies (stochastic or deterministic) as $\mathcal{P}$, the probability law induced by a policy $\mu \in \mathcal{P}$ as $p_\mu$, and the expectation corresponding to $p_\mu$ as $E_\mu [\cdot]$.
% The tuple $(X_k, Y_k, U_k)$ constitutes a controlled hidden Markov model (HMM) \cite{Krishnamurthy2016}.

In general, the controls $U_k$ affect both the state values and the uncertainty associated with them in a phenomenon known as the dual-control effect \cite{Bar1974}.
The dual-control effect is often exploited to solve problems that involve selecting controls for the dual purpose of optimising both a \emph{system-performance measure} dependent on the state and control values (e.g. state and control costs) and a \emph{state-uncertainty measure} dependent on the uncertainty associated with the states (e.g. statistics of state estimates).
%We aim to exploit this phenomenon to pose and \emph{active estimation} or \emph{active obfuscation problems that  either find controls that reduce the uncertainty associated with the states to aid in their estimation, or to find controls that increase the uncertainty associated with the states to hinder (and ideally prevent) their estimation.
%for both active estimation and obfuscation by selecting controls to optimise both a \emph{system-performance measure} dependent on the state and control values (e.g.\ state and control costs), as well as a \emph{state-uncertainty measure} dependent on the uncertainty associated with the states (e.g.\ statistics of state estimates).
%For example, in controlled sensing and robotics, HMMs with their controls selected to simultaneously minimise both costs associated with sensing (such as energy consumption) and measures of state (or estimation) uncertainty (such as the entropy of Bayesian state estimates) have been widely used (cf.~\cite{Krishnamurthy2007,Krishnamurthy2016,Araya2010,Thrun2005, Mu2016}).
As a system-performance measure, we consider the standard additive cost functional
\begin{align}
    \label{eq:costFunctional}
    J(\mu)
    &\triangleq E_\mu \left[ c_T(x_T) + \sum_{k = 0}^{T-1} c_k \left(x_k, u_k \right) \right]
\end{align}
with arbitrary cost functions $c_k : \mathcal{X} \times \mathcal{U} \rightarrow \mathbb{R}$ for $0 \leq k < T$ and $c_T : \mathcal{X} \rightarrow \mathbb{R}$.
As a state-uncertainty measure, we consider the conditional entropy of the state trajectory $X^T$ given measurements $Y^T$ and controls $U^{T-1}$ for $T \geq 0$, i.e.,
\begin{align}
    \label{eq:condEntCriteria}
    H_\mu(X^T | Y^T, U^{T-1}) 
    &\triangleq E_\mu [ H(X^T | y^T, u^{T-1})]
\end{align}
where $H_\mu(X^0 | Y^0, U^{-1}) \triangleq H(X_0 | Y_0)$ does not dependent on the controls (and policy $\mu$), and $H(X^T | y^T, u^{T-1})$ is the pointwise entropy of the conditional pmf $p(x^T | y^T, u^{T-1})$ (which also does not depended on the policy given the controls).
We shall refer to \eqref{eq:condEntCriteria} as the \emph{smoother entropy}.

Our consideration of the smoother entropy \eqref{eq:condEntCriteria} as a state-uncertainty measure is motivated by its relationship to the conditional pmf $p(x^T | y^T, u^{T-1})$, which is the (joint) posterior distribution of concern in Bayesian state estimation --- with Bayesian smoothers computing its marginals $p(x_k | y^T, u^{T-1})$ for $0 \leq k \leq T$ and the Viterbi algorithm computing its mode (cf.\ \cite{Briers2010,Krishnamurthy2016}).
Intuitively, {the smoother entropy describes the uncertainty about the value of the state trajectory $X^T$ given the measurement and control trajectories $Y^T$ and $U^{T-1}$.}
Hence, the smaller (greater) the smoother entropy, the less (more) uncertain we expect state trajectory estimates from smoother-like algorithms.
In the extreme case $H_\mu(X^T | Y^T, U^{T-1}) = 0$, the state trajectory can be uniquely recovered from the measurement and control trajectories.

% We investigate the selection of controls to aid or hinder the estimation of the state trajectory with smoother-like algorithms through optimisation of the smoother entropy.
We seek to find control policies that minimise (arbitrary) combinations of the system-performance measure \eqref{eq:costFunctional} and the smoother entropy \eqref{eq:condEntCriteria} by solving
\begin{align}
    \label{eq:activeEstimation}
    \begin{aligned}
    \inf_{\mu \in \mathcal{P}} \quad & \beta H_\mu (X^T | Y^T, U^{T-1}) + J(\mu)\\
    \mathrm{s.t.} 
    \quad & X_{k+1} | X_k, U_k \sim A^{x_{k+1},x_k}(u_k), \; X_0 \sim \rho\\
    \quad & Y_{k+1} | X_{k+1}, U_k \sim B^{x_{k+1},y_{k+1}}(u_k), \; Y_0 | X_0 \sim B^{x_0,y_0}\\
    \quad & U_k | I_k \sim \mu_k^{i_k}(u_k)
    \end{aligned}
\end{align}
for any given (potentially negative) real-valued constant $\beta \in (-\infty, \infty)$.
When $\beta > 0$, \eqref{eq:activeEstimation} is a formulation of \emph{active state estimation} concerned with finding control policies that minimise the smoother entropy so as to aid the estimation of the state trajectory $X^T$.
When $\beta < 0$, \eqref{eq:activeEstimation} is a novel formulation of \emph{active state obfuscation} concerned with finding control policies that maximise the smoother entropy so as to hinder (and ideally prevent) the estimation of the state trajectory $X^T$.
When $\beta = 0$, \eqref{eq:activeEstimation} reduces to a standard POMDP concerned only with optimising the system-performance measure $J(\mu)$.
The sign of $\beta$ in \eqref{eq:activeEstimation} thus determines whether \eqref{eq:activeEstimation} is an active state estimation or active state obfuscation problem.
The magnitude of $\beta$ in \eqref{eq:activeEstimation} determines the degree to which we are willing to prioritise optimising the smoother entropy (and hence the objectives of active state estimation or obfuscation) over minimising $J(\mu)$.
We will provide further interpretations of \eqref{eq:activeEstimation} in Section \ref{sec:operational}, after we have examined its solution.

We note that \eqref{eq:activeEstimation} resembles entropy-regularised POMDPs and MDPs that have recently appeared in the reinforcement-learning literature (e.g.\ \cite{Lee2021, Haarnoja2017, Haarnoja2018}).
However, the entropy of interest in these reinforcement-learning works is that of the controls or policy, i.e. $H_\mu(U_k | i_k) = -\sum_{u \in \mathcal{U}} \mu_k^{i_k}(u) \log \mu_k^{i_k}(u)$.
In contrast, for active state estimation or obfuscation, the entropies of interest are primarily related to \emph{state distributions}, such as $p_\mu(x_k | y^k, u^{k-1})$ (cf.~\cite[Chapter 8]{Krishnamurthy2016}, \cite{Araya2010, Fehr2018} and references therein), with the novelty of \eqref{eq:activeEstimation} being consideration of the smoother entropy $H_\mu(X^T | Y^T, U^{T-1})$ for both active state estimation and obfuscation (i.e.~$\beta$ positive or negative).

\subsection{POMDP Solution Approach}
To examine the solution of \eqref{eq:activeEstimation} for any $\beta \in (-\infty, \infty)$, let us define the \emph{belief state} $\pi_{k} \in \Delta$ as the conditional pmf of the state $X_k$ given the information state $i_k = (y^k, u^{k-1})$, that is, $\pi_{k}(x) \triangleq p(X_{k} = x | y^k, u^{k-1})$ for $x \in \mathcal{X}$.
The belief state evolves via the Bayesian filter:
\begin{align}
    \label{eq:bayesTemp}
    \pi_{k+1}(x)%\Pi(\pi_{k-1}, y_k, u_{k-1})
    &= \dfrac{ B^{x,y_{k+1}}(u_{k}) \sum_{\bar{x} \in \mathcal{X}} \bar{\pi}_{k+1 | k} (x,\bar{x})}{\sum_{\tilde{x}, \bar{x} \in \mathcal{X}} B^{\tilde{x},y_{k+1}}(u_{k}) \bar{\pi}_{k+1 | k} (\tilde{x},\bar{x})}
\end{align}
for $k \geq 0$ and all $x \in \mathcal{X}$ where $\bar{\pi}_{k+1 | k}(x,\bar{x}) \triangleq p(X_{k+1} = x, X_{k} = \bar{x} | y^{k}, u^{k})$ is the \emph{joint predicted belief state} given by
\begin{align}\label{eq:bayesianPred}
    \bar{\pi}_{k+1 | k}(x,\bar{x})
    &= A^{x,\bar{x}}(u_{k}) \pi_{k}(\bar{x})
\end{align}
for $x,\bar{x} \in \mathcal{X}$.
The Bayesian filter \eqref{eq:bayesTemp} is a mapping of $\pi_k$, $u_k$ and $y_{k+1}$ to $\pi_{k+1}$ that we shall write compactly as
\begin{align}
    \label{eq:bayesianFilter}
    \pi_{k+1}
    &= \Pi(\pi_{k}, u_{k}, y_{k+1})
\end{align}
for $k \geq 0$, with the initial belief state $\pi_0$ given by the mapping $\pi_0(x_0) = B^{x_0,y_0}\rho(x_0) / (\sum_{x \in \mathcal{X}} B^{x,y_0}\rho(x))$ for $x_0 \in \mathcal{X}$, which we shall write as $\pi_0 = \Pi_0(\rho, y_0)$.

When $\beta = 0$, \eqref{eq:activeEstimation} reduces to a standard POMDP so can be reformulated (cf.~\cite[Chapter 7]{Krishnamurthy2016}) as the belief-state MDP
\begin{align}
\label{eq:standardPOMDP}
\begin{aligned}
&\inf_{\bar{\mu}} & & E_{\bar{\mu}} \left[ C_T(\pi_T) + \sum_{k = 0}^{T-1} C_k \left( \pi_k, u_{k} \right) \right]\\ 
&\mathrm{s.t.} & &  \pi_{k+1} = \Pi\left( \pi_{k}, u_{k}, y_{k+1} \right), \; \pi_0 = \Pi_0(\rho, y_0)\\
& & & Y_{k+1} | \pi_k, U_k \sim p(y_{k+1} | \pi_{k}, u_{k}), \; Y_0 | \rho \sim p(y_0 | \rho)\\
& & & U_k = \bar{\mu}_k(\pi_k) \in \mathcal{U}
\end{aligned}
\end{align}
with the optimisation over deterministic policies $\bar{\mu} \triangleq \{\bar{\mu}_k : 0 \leq k < T\}$ defined by functions of the belief state, $\bar{\mu}_k : \Delta \rightarrow \mathcal{U}$.
The cost functions are
$
	C_T (\pi_T)
	\triangleq E_{X_T} [c_T(x_T) | \pi_T]
$
and $
	C_k (\pi_k, u_k)
	\triangleq E_{X_k} [c_k(x_k, u_k) | \pi_k, u_k],
$
with
\begin{align}
    \label{eq:Y_belief}
        p(y_{k+1} | \pi_{k}, u_{k})
        = \sum_{x,\bar{x} \in \mathcal{X}} B^{x,y_{k+1}}(u_{k}) A^{x,\bar{x}}(u_{k}) \pi_{k}(\bar{x})
\end{align}
for $k \geq 0$ and $p(y_0 | \rho) = \sum_{x_0 \in \mathcal{X}} \rho(x_0) B^{x_0,y_0}$.

Numerous techniques based on dynamic programming exist for finding (approximate) solutions to POMDPs of the form of \eqref{eq:activeEstimation} with $\beta = 0$ (and their belief-state MDPs, as shown in \eqref{eq:standardPOMDP}).
These techniques are increasingly able to handle large state, measurement, and control spaces by exploiting structural properties of the cost functions $C_k (\pi_k, u_k)$ and $C_T(\pi_T)$ and the resulting dynamic programming value (or cost-to-go) functions (see \cite{Haugh2020,Walraven2019,Krishnamurthy2016,Araya2010,Garg2019} and references therein).
In particular, the vast majority of POMDP techniques exploit the fact that the cost and value functions of standard POMDPs of the form in \eqref{eq:standardPOMDP} are concave (or PWLC) in the belief state $\pi_k$ for all $u_k \in \mathcal{U}$ (cf.\ \cite{Araya2010} and \cite[Chapter 8.4.4]{Krishnamurthy2016}).\footnote{Due to the control space $\mathcal{U}$ being finite, standard POMDP techniques are not usually concerned with structural properties with respect to the controls.}

However, when $\beta \neq 0$, the presence of the smoother entropy $H_\mu(X^T | Y^T, U^{T-1})$ in \eqref{eq:activeEstimation} complicates its solution in the same manner as standard POMDPs of the form in \eqref{eq:standardPOMDP} with cost and value functions that are additive and concave in the belief state.
Indeed, the smoother entropy has previously been dismissed as difficult or problematic to minimise, due to the correlations between successive states that it captures \cite{Stachniss2005,Valencia2012,Valencia2018}, and the closest (exact) results in \cite{Hernando2005} establish only an additive (non-belief-state) expression for the \emph{pointwise} conditional entropy $H(X^T|y^T)$ for (uncontrolled) hidden Markov models.
In this paper, we therefore focus on establishing novel belief-state forms of the smoother entropy that possess an additive structure that allows us to reformulate and solve \eqref{eq:activeEstimation} using standard techniques regardless of whether $\beta > 0$ or $\beta < 0$.

\section{Additive and Belief-State Forms of the Smoother Entropy}
\label{sec:directedInformation}

In this section, we establish novel additive and belief-state forms of the smoother entropy $H_\mu(X^T | Y^T, U^{T-1})$ using concepts from the Marko-Massey theory of \emph{directed information} \cite{Marko1973,Massey1990,Kramer1998,Massey2005}.
These novel forms will enable us to later reformulate \eqref{eq:activeEstimation} as multiple (fully-observed) belief-state MDPs.

\subsection{Marko-Massey Directed-Information Forms}
To establish our first main result, let us define the \emph{causally conditioned directed information} from the states $X^T$ to the measurements $Y^T$ given the controls $U^{T-1}$ under a policy $\mu \in \mathcal{P}$ as \cite{Kramer1998,Massey2005}
\begin{align}
    \label{eq:causalDirectedInformation}
    I_\mu(X^T \to Y^T \| U^{T-1})
    &\triangleq \sum_{k = 0}^T I_\mu(X^k; Y_k | Y^{k-1}, U^{k-1})
\end{align}
where $I_\mu(X^0; Y_0 | Y^{-1}, U^{-1}) \triangleq I(X_0; Y_0)$, which does not depend on the controls (and hence policy).
Similarly, let the \emph{causally conditioned entropy} of the states $X^T$ given the measurements $Y^{T-1}$ and controls $U^{T-1}$ under $\mu$ be \cite{Kramer1998,Massey2005}
\begin{align}
    \label{eq:causalEntropy}
    H_\mu(X^T \| Y^{T-1}, U^{T-1})
    &\triangleq \sum_{k = 0}^T H_\mu(X_k | X^{k-1}, Y^{k-1}, U^{k-1})
\end{align}
where $H_\mu(X_0 | X^{-1}, Y^{-1}, U^{-1}) \triangleq H(X_0)$, which does not depend on the controls (or policy).

Intuitively, $I_\mu(X^T \to Y^T \| U^{T-1})$ describes the total ``new'' information causally gained over each time-step about the states from the measurements given the controls, whilst $H_\mu(X^T \| Y^{T-1}, U^{T-1})$ describes the total uncertainty about the state trajectory over each time-step given causal knowledge of past states, measurements, and controls.
The following theorem establishes that the (non-causal) smoother entropy $H_\mu(X^T | Y^T, U^{T-1})$ is the difference between $H_\mu(X^T \| Y^{T-1}, U^{T-1})$ and $I_\mu(X^T \to Y^T \| U^{T-1})$.

\begin{theorem}
 \label{theorem:directedInformation}
 Consider any (potentially stochastic) policy $\mu \in \mathcal{P}$.
 Then for any $T \geq 0$,
  \begin{align}
    \label{eq:stageAdditiveCausal}
    \begin{split}
        &H_\mu(X^T | Y^T, U^{T-1})\\
        &\;= H_\mu(X^T \| Y^{T-1}, U^{T-1}) - I_\mu(X^T \to Y^T \| U^{T-1}).
    \end{split}
 \end{align}
\end{theorem}
\begin{IEEEproof}
  We prove \eqref{eq:stageAdditiveCausal} via induction on $T$.
  For $T = 0$,
  \begin{align*}
      &H_\mu(X^0 \| Y^{-1}, U^{-1}) - I_\mu(X^0 \to Y^0 \| U^{-1})\\
      &\quad= H(X_0) - I(X_0; Y_0) = H(X_0 | Y_0)
  \end{align*}
  and so \eqref{eq:stageAdditiveCausal} holds for $T = 0$.
  Suppose then that \eqref{eq:stageAdditiveCausal} holds for trajectory lengths smaller than $T$ where $T > 0$.
  From the definitions of the causally conditioned directed information \eqref{eq:causalDirectedInformation} and causal conditional entropy \eqref{eq:causalEntropy}, we have that
  \begin{align*}
      &I_\mu(X^T \to Y^{T} \| U^{T-1})\\
      &\;= I_\mu(X^{T-1} \to Y^{T-1} \| U^{T-2}) + I_\mu(X^T ; Y_T | Y^{T-1}, U^{T-1})
  \end{align*}
  and
  \begin{align*}
      H_\mu(X^T \| Y^{T-1}, U^{T-1})
      &= H_\mu(X^{T-1} \| Y^{T-2}, U^{T-2})\\
      &\quad+ H_\mu(X_T | X^{T-1}, Y^{T-1}, U^{T-1}).
  \end{align*}
  Combining these two equations gives
  \begin{align}\notag
      &H_\mu(X^T \| Y^{T-1}, U^{T-1}) - I_\mu(X^T \to Y^{T} \| U^{T-1})\\\notag
      &=\hspace{-1.5pt} H_\mu(X^{T-1} \| Y^{T-2}, U^{T-2}) + H_\mu(X_T | X^{T-1}, Y^{T-1}, U^{T-1}) \\\notag
      &\quad\hspace{-1.5pt}- I_\mu(X^{T-1} \to Y^{T-1} \| U^{T-2})
      - I_\mu(X^T ; Y_T | Y^{T-1}, U^{T-1})\\\notag
      &=\hspace{-1.5pt} H_\mu(X^{T-1} | Y^{T-1}, U^{T-2}) + H_\mu(X_T | X^{T-1}, Y^{T-1}, U^{T-1}) \\\label{eq:induct1}
      &\quad\hspace{-1.5pt}- I_\mu(X^T ; Y_T | Y^{T-1}, U^{T-1})
  \end{align}
  where the last equality follows from the induction hypothesis that \eqref{eq:stageAdditiveCausal} holds for trajectories shorter than $T > 0$.
  To simplify \eqref{eq:induct1}, note that the definition of mutual information implies that
  \begin{align}\notag
      &I_\mu(X^T ; Y_T | Y^{T-1}, U^{T-1})\\\notag
      &= H_\mu(X^T | Y^{T-1}, U^{T-1}) - H_\mu (X^T | Y^T, U^{T-1})\\\notag
      &= H_\mu(X^{T-1} | Y^{T-1}, U^{T-2}) + H_\mu(X_T | X^{T-1}, Y^{T-1}, U^{T-1})\\ \label{eq:condMut1}
      &\quad- H_\mu (X^T | Y^T, U^{T-1})
  \end{align}
  where the last equality follows from the chain rule for conditional entropy, and by noting that $U_{T-1}$ is conditionally independent of $X^{T-1}$ given $U^{T-2}$ and $Y^{T-1}$ by virtue of the measurement kernel \eqref{eq:obsProcess} and the feedback control policy $\mu$.
  Substituting \eqref{eq:condMut1} into \eqref{eq:induct1} then gives that
  \begin{align*}
    &H_\mu(X^T \| Y^{T-1}, U^{T-1}) - I_\mu(X^T \to Y^{T} \| U^{T-1})\\
    &\quad= H_\mu(X^T | Y^T, U^{T-1})
  \end{align*}
  and so \eqref{eq:stageAdditiveCausal} holds for $T > 0$. 
  The proof is complete.
\end{IEEEproof}

The causal conditioning on $Y^{T-1}$ in $H_\mu(X^T \| Y^{T-1}, U^{T-1})$ can be omitted in \eqref{eq:stageAdditiveCausal} since the Markov property of the state process $X_k$ and \eqref{eq:causalEntropy} implies that
$
  H_\mu(X^T \| Y^{T-1}, U^{T-1})
%    &= \sum_{k = 1}^T H(X_k | X^{k-1}, Y^{k-1}, U^{k-1})\\
%    &= \sum_{k = 1}^T H(X_k | X^{k-1}, U^{k-1})\\
    = H_\mu(X^T \| U^{T-1}).
$
Hence, \eqref{eq:stageAdditiveCausal} resembles the trivial expression of the smoother entropy as
\begin{align}
	\label{eq:standardTrajectroyForm}
	\begin{split}
    &H_\mu(X^T|Y^T,U^{T-1})\\
    &\quad= H_\mu(X^T | U^{T-1}) - I_\mu(X^T; Y^T | U^{T-1}).
   \end{split}
\end{align}
Expressions \eqref{eq:stageAdditiveCausal} and \eqref{eq:standardTrajectroyForm} are subtly different since the causally conditioned directed information and entropy terms in \eqref{eq:stageAdditiveCausal} involve conditional probabilities of the states $X_k$ given only the information state $I_k = (Y^k, U^{k-1})$, whilst the standard conditional entropy and mutual information terms in \eqref{eq:standardTrajectroyForm} involve conditional probabilities of the states $X_k$ given the entire trajectories of measurements $Y^{T}$ and controls $U^{T-1}$. 
This difference means that \eqref{eq:stageAdditiveCausal} will lead directly to belief-state forms of the smoother entropy.

To express the smoother entropy in terms of the belief state, we require the following corollary to Theorem \ref{theorem:directedInformation}.

\begin{corollary}
\label{corollary:additive}
Consider any (potentially stochastic) policy $\mu \in \mathcal{P}$.
 The smoother entropy has the additive forms:
 \begin{align}
 \notag
        &H_\mu(X^T | Y^T, U^{T-1})\\\label{eq:firstAdditiveForm}
        &= \sum_{k = 0}^T [H_\mu(X_k | X_{k-1}, U_{k-1}) - I_\mu(X_k ; Y_k | Y^{k-1}, U^{k-1})]\\ \label{eq:secondAdditiveForm}
%        &\quad= E_\mu \Bigg[ \sum_{k = 1}^T \big[ H(X_k | y^k, u^{k-1}) - H(X_k | y^{k-1}, u^{k-1})\\
         %&\qquad\qquad+ H(X_k | X_{k-1}, y^{k-1}, u^{k-1}) \big] \Bigg].         
       &= \sum_{k = 0}^T [H_\mu(X_k | Y^k, U^{k-1}) - I_\mu(X_k; X_{k-1} | Y^{k-1}, U^{k-1})]\\\label{eq:thirdAdditiveForm}
       &= H_\mu(X_T | Y^{T}, U^{T-1}) + \sum_{k = 0}^{T-1} H_\mu(X_{k} | X_{k+1}, Y^k, U^k)
\end{align}
where $H_\mu(X_0 | X_{-1}, U_{-1}) \triangleq H(X_0)$, $I_\mu(X_0; Y_0 | Y^{-1}, U^{-1}) \triangleq I(X_0;Y_0)$, $H_\mu(X_0 | Y^0, U^{-1}) \triangleq H(X_0 | Y_0)$, and $I_\mu(X_0; X_{-1} | Y^0, U^0) \triangleq 0$.
\end{corollary}
\begin{IEEEproof}
  The definition of mutual information implies
  \begin{align*}
      &I_\mu(X^k; Y_k | Y^{k-1}, U^{k-1})\\
      &\quad= H_\mu(Y_k | Y^{k-1}, U^{k-1}) - H_\mu(Y_k | X^k, Y^{k-1}, U^{k-1})\\
      &\quad= H_\mu(Y_k | Y^{k-1}, U^{k-1}) - H_\mu(Y_k | X_k, Y^{k-1}, U^{k-1})\\
      &\quad= I_\mu(X_k; Y_k | Y^{k-1}, U^{k-1})
  \end{align*}
  where the second equality holds due to the Markov property of the state process $X_k$.
  Thus, \eqref{eq:causalDirectedInformation} is equivalent to
  \begin{align*}
    I_\mu(X^T \to Y^T \| U^{T-1})
    = \sum_{k = 0}^T I_\mu(X_k; Y_k | Y^{k-1}, U^{k-1})
    %&\quad= \sum_{k = 1}^T \left[ H(X_k | Y^{k-1}, U^{k-1})  - H(X_k | Y^{t}, U^{k-1}) \right].
    \end{align*}
    with $I_\mu(X_0; Y_0 | Y^{-1}, U^{-1}) = I(X_0; Y_0)$.
    Substituting this expression and the definition of the causally conditioned entropy \eqref{eq:causalEntropy} into \eqref{eq:stageAdditiveCausal}, noting also that
    \begin{align*}
    	H_\mu(X_k | X^{k-1}, Y^{k-1}, U^{k-1}) = H_\mu(X_k | X_{k-1}, U_{k-1})
    \end{align*}
    due to the Markov property of the state $X_k$, gives \eqref{eq:firstAdditiveForm}.
    
    Now, the summands in \eqref{eq:firstAdditiveForm} can be rewritten as
      \begin{align*}
      &H_\mu(X_k | X_{k-1}, U_{k-1}) - I_\mu(X_k; Y_k | Y^{k-1}, U^{k-1})\\
      &\;= H_\mu(X_k | X_{k-1}, Y^{k-1}, U^{k-1}) - I_\mu(X_k; Y_k | Y^{k-1}, U^{k-1})\\
      &\;= H_\mu(X_k | X_{k-1}, Y^{k-1}, U^{k-1}) - H_\mu(X_k | Y^{k-1}, U^{k-1})\\
      &\qquad+ H_\mu(X_k | Y^{k}, U^{k-1})\\
      &\;= H_\mu(X_k | Y^{k}, U^{k-1}) - I_\mu(X_k; X_{k-1} | Y^{k-1}, U^{k-1})
    \end{align*}
    where the first equality holds due to the Markov property of the state $X_k$, and the remainder follow from the definitions of the conditional mutual informations between $X_k$ and $Y_k$, and $X_k$ and $X_{k-1}$.
    The second additive form \eqref{eq:secondAdditiveForm} follows.
    
    Finally, symmetry of the mutual information in \eqref{eq:secondAdditiveForm} implies
   \begin{align*}
    &H_\mu(X^T | Y^T, U^{T-1})\\
    &\;= \sum_{k = 0}^T [ H_\mu(X_k | Y^k, U^{k-1}) - I_\mu(X_k; X_{k-1} | Y^{k-1}, U^{k-1})]\\
    &\;= \sum_{k = 0}^T [ H_\mu(X_k | Y^{k}, U^{k-1}) - H_\mu(X_{k-1} | Y^{k-1}, U^{k-1})\\
    &\qquad+ H_\mu(X_{k-1} | X_k, Y^{k-1}, U^{k-1})]\\
    &\;= H_\mu(X_T | Y^{T}, U^{T-1}) + \sum_{k = 1}^{T} H_\mu(X_{k-1} | X_{k}, Y^{k-1}, U^{k-1})
	\end{align*}
	where the last equality follows by noting that consecutive entropy terms $H_\mu(X_k | Y^k, U^{k-1})$ cancel since $H_\mu(X_{k-1} | Y^{k-1}, U^{k-1}) = H_\mu(X_{k-1} | Y^{k-1}, U^{k-2})$ by virtue of the state $X_{k-1}$ being conditionally independent of the control $U_{k-1}$ given $Y^{k-1}$ and $U^{k-2}$ due to \eqref{eq:stateProcess} and the feedback policy (cf.\ the conditions of Theorem \ref{theorem:directedInformation}).
	The third additive form \eqref{eq:thirdAdditiveForm} follows and the proof is complete.  
%    \begin{align*}
%        &H(X^T | Y^T, U^{T-1})\\
%        &\quad= \sum_{k = 0}^T \big[ H(X_k | Y^k, U^{k-1}) - H(X_k | Y^{k-1}, U^{k-1}) \\
%        &\qquad \qquad+ H(X_k | X_{k-1}, Y^{k-1}, U^{k-1}) \big].
%    \end{align*}
%    The corollary result follows by recalling that the conditional entropies in this expression can be written as the expectation of pointwise conditional entropies.
\end{IEEEproof}

The additive forms established in Corollary \ref{corollary:additive} each provide different interpretations of the smoother entropy.
The first form \eqref{eq:firstAdditiveForm} provides the interpretation of the smoother entropy as the sum of the uncertainty from the state transitions, i.e.\ $H_\mu(X_k | X_{k-1}, U_{k-1})$, minus the information about the states gained from the measurements, i.e.\ $I_\mu(X_k;Y_k | Y^{k-1}, U^{k-1})$.
The second form \eqref{eq:secondAdditiveForm} suggests that the smoother entropy can be viewed as the sum of the marginal (or instantaneous) state uncertainties, i.e.\ $H_\mu(X_k | Y^{k}, U^{k-1})$, minus the dependency between consecutive states, i.e.\ $I_\mu(X_k;X_{k+1}|Y^k, U^k)$.
Finally, the third form \eqref{eq:thirdAdditiveForm} offers an interpretation of the smoother entropy backwards in time, with it being the uncertainty associated with the final state $X_T$, i.e., $H_\mu(X_T | Y^T, U^{T-1})$, plus the uncertainty accumulated via (backwards) state transitions, i.e., $H_\mu(X_k | X_{k+1}, Y^{k}, U^{k})$.

\subsection{Belief-State Forms of the Smoother Entropy}
The significance of the forms of the smoother entropy established in Corollary \ref{corollary:additive} is that they lead to expressions of it in terms of the belief state $\pi_k$, as we shall now show.

\subsubsection{First Belief-State Form}

The third additive form of the smoother entropy established in Corollary \ref{corollary:additive}, i.e.\ \eqref{eq:thirdAdditiveForm}, can be expressed in terms of pointwise entropies as
\begin{align*}
 &H_\mu(X^T | Y^T, U^{T-1})\\
 &= E_\mu \left[ H(X_T | y^{T}, u^{T-1}) + \sum_{k = 0}^{T-1} H(X_{k} | X_{k+1}, y^{k}, u^{k}) \right].
\end{align*}
Since $H(X_T | y^{T}, u^{T-1})$ is the entropy of the terminal belief state $\pi_T$, it is solely a function of $\pi_T$ in the sense that
\begin{align}\notag
    H(X_T | y^T, u^{T-1})
    &= - \sum_{x \in \mathcal{X}} \pi_T(x) \log \pi_T(x)\\\label{eq:second_terminal_cost_entropy}
    &\triangleq \tilde{g}_T (\pi_T).
\end{align}
Similarly, the conditional entropy $H(X_{k} | X_{k+1}, y^{k}, u^{k})$ is a function of $\pi_k$ and $u_k$ due to it being defined in terms of the joint pmf $p(x_k, x_{k+1} | y^k, u^k)$ (which is the joint predicted belief $\bar{\pi}_{k+1 | k}$ in \eqref{eq:bayesianPred}) and the conditional pmf $p(x_k | x_{k+1}, y^k, u^k)$ (which can be computed from the joint predicted belief $\bar{\pi}_{k+1 | k}$ via appropriate marginalisation and division).
Hence,
\begin{align}\notag
    &H(X_{k} | X_{k+1}, y^{k}, u^{k})\\\notag
%    &= - \sum_{i,j = 1}^{N_x} \bar{\pi}_{k+1 | k}(i, j) \log \dfrac{\bar{\pi}_{k+1 | k}(i, j)}{\sum_{m = 1}^{N_x} \bar{\pi}_{k+1 | k}(i, m)}\\\notag
    &= - \sum_{x,\bar{x} \in \mathcal{X}} A^{x,\bar{x}}(u_{k}) \pi_{k}(\bar{x}) \log \dfrac{A^{x,\bar{x}}(u_{k}) \pi_{k}(\bar{x})}{ \sum_{\tilde{x} \in \mathcal{X}} A^{x,\tilde{x}}(u_{k}) \pi_{k}(\tilde{x})}\\\label{eq:second_running_cost_entropy}
    &\triangleq \tilde{g}(\pi_k, u_k).
\end{align}
Thus, \eqref{eq:thirdAdditiveForm} yields the belief-state form of the smoother entropy:
\begin{align}\label{eq:secondBeliefAdditiveForm}
 H_\mu(X^T | Y^T, U^{T-1})
 = E_\mu \left[ \tilde{g}_T(\pi_T) + \sum_{k = 0}^{T-1} \tilde{g} (\pi_{k}, u_{k}) \right].
\end{align}

\subsubsection{Second Belief-State Form}
The second additive form in Corollary \ref{corollary:additive}, i.e.\ \eqref{eq:secondAdditiveForm}, yields an alternative belief-state expression for the smoother entropy.
Specifically, by recalling the definition of mutual information, \eqref{eq:secondAdditiveForm} can be expressed as the expectation of the sum of pointwise entropies, namely,
\begin{align}\notag
 &H_\mu(X^T | Y^T, U^{T-1})\\\notag
 &\;= H(X_0 | Y_0) + E_\mu \Bigg[ \sum_{k = 0}^{T-1} \big[H(X_{k+1} | y^{k+1}, u^{k}) \\ \label{eq:firstBeliefAdditiveFormPointWise}
 &\qquad - H(X_{k+1} | y^{k}, u^{k}) + H(X_{k+1} | X_{k}, y^{k}, u^{k}) \big]\Bigg].
\end{align}

The first term in \eqref{eq:firstBeliefAdditiveFormPointWise}, $H(X_0 | Y_0)$, is the conditional entropy of the initial state $X_0$ given the initial observation $Y_0$, which depends only on the initial state pmf $\rho$ and $B^{x_0,y_0}$ via $p(x_0,y_0) = B^{x_0,y_0}\rho(x_0)$, and not on the controls $U^{T-1}$ or policy, $\mu$.
Since this term in uncontrolled (and fully determined by the initial conditions of the problem \eqref{eq:activeEstimation}), we write it outside of the policy-dependent expectation.

Considering the terms in the expectation in \eqref{eq:firstBeliefAdditiveFormPointWise}, the first term, $H(X_{k+1} | y^{k+1}, u^{k})$, is the entropy of $\pi_{k+1}$ given by
\begin{align}\notag
    H(X_{k+1} | y^{k+1}, u^{k})
    &= - \sum_{x \in \mathcal{X}} \pi_{k+1}(x) \log \pi_{k+1}(x)\\\label{eq:first_current_cost_entropy}
    &\triangleq \tilde{\ell}_1 (\pi_{k}, u_{k}, y_{k+1})
\end{align}
where the last line holds since $\pi_{k+1}$, and hence $H(X_{k+1} | y^{k+1}, u^{k})$, is a function, $\tilde{\ell}_1$, of $\pi_{k}$, $y_{k+1}$ and $u_{k}$ via the Bayesian filter \eqref{eq:bayesianFilter}.
Similarly, the second term in the expectation in \eqref{eq:firstBeliefAdditiveFormPointWise} is a function of $\pi_{k}$ and $u_{k}$, namely, 
\begin{align}\notag
    &H(X_{k+1} | y^{k}, u^{k})\\\notag
    &\quad= - \sum_{x,\bar{x} \in \mathcal{X}} A^{x,\bar{x}}(u_{k}) \pi_{k} (\bar{x}) \log \sum_{\tilde{x} \in \mathcal{X}} A^{\bar{x},\tilde{x}}(u_{k}) \pi_{k} (\tilde{x}) \\\label{eq:first_pred_cost_entropy}
    &\quad\triangleq \tilde{\ell}_2 (\pi_{k}, u_{k}).
\end{align}
Finally, the last term in the expectation in \eqref{eq:firstBeliefAdditiveFormPointWise}, $H(X_{k+1} | X_{k}, y^{k}, u^{k})$, is a function of $\pi_{k}$ and $u_{k}$, namely,
\begin{align}\notag
    &H(X_{k+1} | X_{k}, y^{k}, u^{k})\\\notag
    &\quad= - \sum_{x,\bar{x} \in \mathcal{X}} A^{x,\bar{x}}(u_{k}) \pi_{k} (\bar{x}) \log A^{x,\bar{x}}(u_{k})\\\label{eq:first_cond_cost_entropy}
    &\quad\triangleq \tilde{\ell}_3(\pi_{k}, u_{k})
\end{align}
since $p(x_{k+1} | x_k, y^k, u^k) = p(x_{k+1} | x_k, u_k)$ due to the Markov property of the state.
Thus, \eqref{eq:firstBeliefAdditiveFormPointWise} yields the belief-state form of the smoother entropy:
\begin{align}\notag
 &H_\mu(X^T | Y^T, U^{T-1})\\\label{eq:firstBeliefAdditiveForm}
 &\quad= H(X_0 | Y_0) + E_\mu \left[ \sum_{k = 0}^{T-1} \tilde{\ell} (\pi_{k}, u_{k}, y_{k+1}) \right]
\end{align}
where
\begin{align}
    \label{eq:firstBeliefAdditiveFormFunction}
    \begin{split}
        \tilde{\ell} (\pi_{k}, u_{k}, y_{k+1})
        &\triangleq \tilde{\ell}_1 (\pi_{k}, u_{k}, y_{k+1}) - \tilde{\ell}_2 (\pi_{k}, u_{k})\\
        &\quad+ \tilde{\ell}_3 (\pi_{k}, u_{k}).
    \end{split}
\end{align}

% \subsubsection{Comparison of Belief-State Forms}
% Whilst the belief-state forms of the smoother entropy in \eqref{eq:secondBeliefAdditiveForm} and \eqref{eq:firstBeliefAdditiveForm} are equivalent in total, the functions $\tilde{\ell}$ and $\tilde{g}$ possess different properties that may make one form more attractive than the other in specific cases.
% For example, evaluating $\tilde{\ell}$ involves computing three entropy terms whilst evaluating $\tilde{g}$ involves computing a single entropy term, thus evaluating \eqref{eq:secondBeliefAdditiveForm} is less computationally complex than evaluating \eqref{eq:firstBeliefAdditiveForm}.
%Similarly, since \eqref{eq:secondBeliefAdditiveForm} involves the sum of individual conditional entropy terms that are likely to be concave in the individual belief states $\pi_k$, establishing important structural properties of solutions to optimisation problems involving it such as our active estimation problem \eqref{eq:activeEstimation} is likely to be more straightforward than using \eqref{eq:firstBeliefAdditiveForm}.

We shall exploit the belief-state forms of the smoother entropy in \eqref{eq:secondBeliefAdditiveForm} and \eqref{eq:firstBeliefAdditiveForm} to solve \eqref{eq:activeEstimation} for any $\beta \in (-\infty, \infty)$ in the same manner as standard POMDPs.
That is, we shall reformulate \eqref{eq:activeEstimation} for any $\beta \in (-\infty, \infty)$ as a belief-state MDP with cost and value functions that are concave in the belief state.
Surprisingly, we will show that for $\beta > 0$, \eqref{eq:activeEstimation} has concave costs when optimising the belief-state form of the smoother entropy in \eqref{eq:secondBeliefAdditiveForm} but not when optimising that in \eqref{eq:firstBeliefAdditiveForm}, and \emph{vice versa} for $\beta < 0$.

\section{Belief-State MDP Reformulations, Structural Results, and Bounded-Error Solutions}
\label{sec:activeEst}

In this section, we establish two distinct belief-state MDP reformulations of \eqref{eq:activeEstimation} based on the novel belief-state expressions of the smoother entropy in \eqref{eq:secondBeliefAdditiveForm} and \eqref{eq:firstBeliefAdditiveForm}.
We provide dynamic programming descriptions of the value functions and optimal solutions of these belief-state MDPs, including their structural properties.
We then exploit these results to identify bounded-error (approximate) solutions to \eqref{eq:activeEstimation} for any $\beta \in (-\infty, \infty)$ using a standard POMDP solution technique.

\subsection{Belief-State MDP Reformulations}
\label{subsect:beliefMDPs}

The following theorem establishes two distinct belief-state MDP reformulations of \eqref{eq:activeEstimation} using \eqref{eq:secondBeliefAdditiveForm} and \eqref{eq:firstBeliefAdditiveForm}.

\begin{theorem}
\label{theorem:activeEstimationMDP}
Consider \eqref{eq:activeEstimation} with any $\beta \in (-\infty, \infty)$.
Define
\begin{align*}
    g_k^\beta(\pi_{k}, u_{k})
    &\triangleq E_{X_{k}} \left[ \beta \tilde{g}(\pi_k, u_k) + c_k (x_{k}, u_{k}) | \pi_{k}, u_{k} \right]
\end{align*}
and
\begin{align*}
    &\ell_k^\beta(\pi_{k}, u_{k})\\
    &\quad\triangleq E_{Y_{k+1}, X_{k}} \left[ \left. \beta \tilde{\ell} \left( \pi_{k}, u_{k}, y_{k+1}\right) + c_k (x_{k}, u_{k}) \right| \pi_{k}, u_{k} \right]
\end{align*}
for $0 \leq k < T$ with
$
    g_T^\beta(\pi_{T})
    \triangleq E_{X_{T}} \left[ \beta \tilde{g}_T(\pi_T) + c_T (x_{T}) | \pi_{T} \right]
$
and
$
    \ell_T(\pi_{T})
    \triangleq E_{X_{T}} \left[ \left. c_T (x_{T}) \right| \pi_{T} \right]
$.
Then, \eqref{eq:activeEstimation} is equivalent to the belief-state MDP
\begin{align}
\label{eq:secondActiveEstimationMDP}
\begin{aligned}
\inf_{\bar{\mu}} E_{\bar{\mu}} \left[ g_T^\beta(\pi_T) + \sum_{k = 0}^{T-1} g_k^\beta \left( \pi_{k}, u_{k} \right) \right],
\end{aligned}
\end{align}
and also to the second belief-state MDP
\begin{align}
\label{eq:firstActiveEstimationMDP}
\begin{aligned}
\beta H(X_0 | Y_0) + \inf_{\bar{\mu}} E_{\bar{\mu}} \left[ \ell_T(\pi_{T}) + \sum_{k = 0}^{T-1} \ell_k^\beta \left( \pi_{k}, u_{k} \right) \right].
% &\mathrm{s.t.} & &  \pi_{k+1} = \Pi\left( \pi_{k}, u_k, y_{k+1} \right)\\
% & & & Y_{k+1} \sim p(y_{k+1} | \pi_{k}, u_{k})\\
% & & & \mathcal{U} \ni U_k \sim \bar{\mu}_k(\pi_k),
\end{aligned}
\end{align}
Both infima are over deterministic belief-state policies $\bar{\mu} = \{\bar{\mu}_k : 0 \leq k < T\}$ with $\bar{\mu}_k : \Delta \rightarrow \mathcal{U}$ being functions of $\pi_k$, and are subject to the constraints
\begin{align}
\label{eq:beliefStateConstraints}
\begin{aligned}
& & & \pi_{k+1} = \Pi\left( \pi_{k}, u_{k}, y_{k+1} \right), \; \pi_0 = \Pi_0(\rho, y_0)\\
& & & Y_{k+1} | \pi_k, U_k \sim p(y_{k+1} | \pi_{k}, u_{k}), \; Y_0 | \rho \sim p(y_0 | \rho)\\
& & & U_k = \bar{\mu}_k(\pi_k) \in \mathcal{U}
\end{aligned}
\end{align}
for $0 \leq k < T$.
\end{theorem}
\begin{IEEEproof}
Substituting \eqref{eq:secondBeliefAdditiveForm} into \eqref{eq:activeEstimation} gives
\begin{align*}
   %&H(X^T | Y^T, U^{T-1}) + E_\mu \left[ c_T(x_T) + \sum_{k = 0}^{T-1} c_k \left(x_k, u_k\right) \right]\\
   E_\mu \left[ c_T(x_T) + \beta \tilde{g}_T(\pi_T) + \sum_{k = 1}^{T-1} \left\{ \beta \tilde{g}(\pi_k, u_k) +  c_k \left(x_k, u_k\right) \right\} \right].
\end{align*}
The linearity and tower properties of expectation imply that
$
    E_\mu \left[ c_T(x_T) + \beta \tilde{g}_T(\pi_T) \right]
    %&= E_{Y^T} \left[ E_{X_T} \left[ c_T(x_T) + \tilde{g}_T(\pi_T) | \pi_T \right] \right]\\
    = E_\mu [ g_T^\beta(\pi_T) ],
$
and,
$
    E_\mu \left[ \beta \tilde{g}(\pi_k, u_k) +  c_k \left(x_k, u_k\right) \right]
    %&= E_{Y^k} \left[ E_{X_k} \left[ \tilde{g}(\pi_k, u_k) +  c_k \left(x_k, u_k\right) | \pi_k, u_k \right] \right]\\
    = E_\mu [ g_k^\beta(\pi_k, u_k) ]
$.
The optimisation objective in \eqref{eq:activeEstimation} is thus equivalently
\begin{align*}
E_\mu \left[ g_T^\beta(\pi_T) + \sum_{k = 0}^{T-1} g_k^\beta(\pi_k, u_k) \right].
\end{align*}
The constraints in \eqref{eq:activeEstimation} imply that $\pi_{k}$ must satisfy the Bayesian filter mapping $\Pi$ with $\pi_0 = \Pi_0(\rho, y_0)$, and that the observations $Y_{k+1}$ given the belief state are distributed according to \eqref{eq:Y_belief} with $Y_0 | \rho \sim p(y_0 | \rho)$.
Thus, \eqref{eq:activeEstimation} is equivalent to:
\begin{align}
\label{eq:beliefStateMDPTemp}
\begin{aligned}
&\inf_{\bar{\mu}} & & E_\mu \left[ g_T^\beta(\pi_T) + \sum_{k = 0}^{T-1} g_k^\beta(\pi_k, u_k) \right]\\ 
&\mathrm{s.t.} & &  \pi_{k+1} = \Pi\left( \pi_{k}, u_{k}, y_{k+1} \right), \; \pi_0 = \Pi_0(\rho, y_0)\\
& & & Y_{k+1} | \pi_k, U_k \sim p(y_{k+1} | \pi_{k}, u_{k}), \; Y_0 | \rho \sim p(y_0 | \rho)\\
& & & U_k | I_k \sim \mu_k^{i_k}(u_k).
\end{aligned}
\end{align}
Furthermore, since the belief state $\pi_k$ is a sufficient statistic for $i_k = (y^k,u^{k-1})$ (cf.\ \cite[Section 5.4.1]{Bertsekas2005}), we can equivalently consider belief-state policies $\mu = \{ \mu_k^{\pi_k} : k \geq 0\}$ with pmfs
\begin{align*}
    \mu_k^{\pi_k}(u_k) 
    \triangleq p(U_k = u_k | \pi_k).
\end{align*}
The constraint $U_k | I_k \sim \mu_k^{i_k}(u_k)$ in \eqref{eq:beliefStateMDPTemp} is thus equivalently $U_k | \pi_k \sim \mu_k^{\pi_k}(u_k)$ (see \cite[Section 5.4.1]{Bertsekas2005} for detailed justification).
It follows that \eqref{eq:beliefStateMDPTemp} is a (fully-observed) MDP with continuous state-space $\Delta$.
Finally, standard MDP results (cf.~\cite[Chapter 4]{Bertsekas2005} and \cite[Section 6.3 and Theorem 6.2.2]{Krishnamurthy2016}) give that there is no loss of optimality in considering only deterministic policies (of the belief state $\pi_k$), proving the equivalence of \eqref{eq:secondActiveEstimationMDP} to \eqref{eq:activeEstimation} under the constraints \eqref{eq:beliefStateConstraints}.

The equivalence of \eqref{eq:firstActiveEstimationMDP} under the constraints \eqref{eq:beliefStateConstraints} to \eqref{eq:activeEstimation} is proved similarly.
Specifically, substituting \eqref{eq:firstBeliefAdditiveForm} into \eqref{eq:activeEstimation} gives
\begin{align*}
   %&H(X^T | Y^T, U^{T-1}) + E_\mu \left[ c_T(x_T) + \sum_{k = 0}^{T-1} c_k \left(x_k, u_k\right) \right]\\
   &\beta H(X_0 | Y_0)\\
   &\quad+ E_\mu \Bigg[ c_T(x_T) + \sum_{k = 0}^{T-1} \big\{ \beta \tilde{\ell}(\pi_k, u_k, y_{k+1}) +  c_k \left(x_k, u_k\right) \big\} \Bigg].
\end{align*}
The linearity and tower properties of expectation imply that
$
    E_\mu \left[ c_T(x_T) \right]
    %&= E_{Y^T} \left[ E_{X_T} \left[ c_T(x_T) + \tilde{g}_T(\pi_T) | \pi_T \right] \right]\\
    = E_\mu [ \ell_T(\pi_T) ],
$
and similarly,
$
    E_\mu [ \beta \tilde{\ell}(\pi_k, u_k, y_{k+1}) +  c_k \left(x_k, u_k\right) ]
    %&= E_{Y^k} \left[ E_{X_k} \left[ \tilde{g}(\pi_k, u_k) +  c_k \left(x_k, u_k\right) | \pi_k, u_k \right] \right]\\
    = E_\mu [ \ell_k^\beta(\pi_k, u_k) ]
$
noting that $\pi_k$ is a deterministic function of $(y^k, u^{k-1})$ via \eqref{eq:bayesianFilter}.
% \begin{align}
%     \label{eq:measurementsState}
%     \begin{split}
%         p(y_{k+1} | \pi_{k}, u_{k})
%         &= \sum_{i,j = 1}^{N_x} B^i(y_{k+1}, u_k) A^{ij}(u_k) \pi_k(j).
%         \end{split}
% \end{align}
Thus, the optimisation objective in \eqref{eq:activeEstimation} is
\begin{align*}
%   &H(X^T | Y^T, U^{T-1}) + E_\mu \left[ c_T(x_T) + \sum_{k = 1}^{T-1} c_k \left(x_k, u_k\right) \right]\\
   \beta H(X_0 | Y_0) + E_\mu \left[ \ell_T(\pi_T) + \sum_{k = 0}^{T-1} \ell_k^\beta(\pi_k, u_k) \right].
\end{align*}
By noting that the first term $\beta H(X_0|Y_0)$ is constant with respect to the controls (and is determined by the constraints on $X_0$ and $Y_0$), we have that \eqref{eq:activeEstimation} is optimised by policies solving
\begin{align*}
\begin{aligned}
&\inf_{\bar{\mu}} & & E_\mu \left[ \ell_T(\pi_T) + \sum_{k = 0}^{T-1} \ell_k^\beta(\pi_k, u_k) \right]\\
&\mathrm{s.t.} & &  \pi_{k+1} = \Pi\left( \pi_{k}, u_{k}, y_{k+1} \right), \; \pi_0 = \Pi_0(\rho, y_0)\\
& & & Y_{k+1} | \pi_k, U_k \sim p(y_{k+1} | \pi_{k}, u_{k}), \; Y_0 | \rho \sim p(y_0 | \rho)\\
& & & U_k | I_k \sim \mu_k^{i_k}(u_k).
\end{aligned}
\end{align*}
As in the case of the first belief-state MDP reformulation \eqref{eq:secondActiveEstimationMDP}, it suffices to consider deterministic belief-state policies in solving this optimisation, and the proof is complete.
\end{IEEEproof}

\begin{remark}
Note that the term $\beta H(X_0 | Y_0)$ in \eqref{eq:firstActiveEstimationMDP} is constant under the constraints \eqref{eq:beliefStateConstraints}, and so does not affect the optimal policy.
Thus, \eqref{eq:firstActiveEstimationMDP} can equivalently be written
\begin{align*}
\begin{aligned}
\inf_{\bar{\mu}} E_{\bar{\mu}} \left[ \ell_T(\pi_{T}) + \sum_{k = 0}^{T-1} \ell_k^\beta \left( \pi_{k}, u_{k} \right) \right].
% &\mathrm{s.t.} & &  \pi_{k+1} = \Pi\left( \pi_{k}, u_k, y_{k+1} \right)\\
% & & & Y_{k+1} \sim p(y_{k+1} | \pi_{k}, u_{k})\\
% & & & \mathcal{U} \ni U_k \sim \bar{\mu}_k(\pi_k),
\end{aligned}
\end{align*}
\end{remark}

We next examine the dynamic programming equations associated with the two belief-state MDP reformulations of our active estimation or obfuscation problem \eqref{eq:activeEstimation} in \eqref{eq:secondActiveEstimationMDP} and \eqref{eq:firstActiveEstimationMDP}.

\subsection{Dynamic Programming Equations}
\label{subsec:dynProgramEst}
%Given the belief-state MDP reformulations of our active estimation or obfuscation problem in Theorem \ref{theorem:activeEstimationMDP}.
%Under deterministic policies, we note that the expectations over $Y^T$ and $U^{T-1}$ in the results of Theorem \ref{theorem:activeEstimationMDP} reduce to expectations over the measurements $Y^T$. 
%, along with the dynamic programming recursions they satisfy and the optimal policies they yield.
The value (or cost-to-go) function of our first belief-state MDP reformulation \eqref{eq:secondActiveEstimationMDP} is defined as
\begin{align*}
    V_k^{\beta,g}(\pi_k)
    \triangleq \inf_{\bar{\mu}_k^{T-1}} E_{\bar{\mu}_k^{T-1}} \left[ \left. g_T^\beta(\pi_T) + \sum_{m = k}^{T-1} g_m^\beta \left( \pi_{m}, u_{m} \right) \right| \pi_{k} \right]
\end{align*}
for $0 \leq k < T$ with $V_T^{\beta,g}(\pi_T) \triangleq g_T^\beta(\pi_T)$ where $\bar{\mu}_k^{T-1}$ denotes the subsequence of functions $\{\bar{\mu}_k, \bar{\mu}_{k+1}, \ldots, \bar{\mu}_{T-1}\}$ from the deterministic belief-state policy $\bar{\mu} = \{\bar{\mu}_0, \bar{\mu}_1, \ldots, \bar{\mu}_{T-1}\}$.
Similarly, the value function of our second belief-state MDP reformulation \eqref{eq:firstActiveEstimationMDP} {(which omits the constant $\beta H(X_0|Y_0)$)} is
\begin{align*}
    V_k^{\beta,\ell}(\pi_k)
    \triangleq \inf_{\bar{\mu}_k^{T-1}} E_{\bar{\mu}_k^{T-1}} \left[ \left. \ell_T(\pi_T) + \sum_{m = k}^{T-1} \ell_m^\beta \left( \pi_{m}, u_{m} \right) \right| \pi_{k} \right]
\end{align*}
for $0 \leq k < T$ and $V_T^{\beta,\ell}(\pi_T) \triangleq \ell_T(\pi_T)$.
By following standard dynamic programming arguments (cf.~\cite[Section 8.4.3]{Krishnamurthy2016}), the value function $V_k^{\beta,g}$ of \eqref{eq:secondActiveEstimationMDP} satisfies
\begin{align}
    \label{eq:secondEstValueFunction}
    \begin{split}
    V_k^{\beta,g}(\pi_k)
    &= \inf_{u_{k} \in \mathcal{U}} \{ g_k^\beta ( \pi_k, u_{k} ) \\
    &\quad + E_{Y_{k+1}} [ V_{k+1}^{\beta,g}(\Pi(\pi_{k}, u_{k}, y_{k+1})) | \pi_k, u_{k} ] \}
    \end{split}
\end{align}
for $0 \leq k < T$ with $V_T^{\beta,g}(\pi_T) = g_T^\beta(\pi_T)$ and where the distribution of $Y_{k+1}$ given $(\pi_k, u_k)$ is given by \eqref{eq:Y_belief}.
Similarly, the value function $V_k^{\beta,\ell}$ of \eqref{eq:firstActiveEstimationMDP} satisfies
\begin{align}
    \label{eq:firstEstValueFunction}
    \begin{split}
    V_k^{\beta,\ell}(\pi_k)
    &= \inf_{u_k \in \mathcal{U}} \{ \ell_k^\beta (\pi_{k}, u_{k})\\
    &\quad+ E_{Y_{k+1}} [ V_{k+1}^{\beta,\ell}(\Pi(\pi_{k}, u_k, y_{k+1})) | \pi_k, u_k ] \}
    \end{split}
\end{align}
for $0 \leq k < T$ with $V_T^{\beta,\ell}(\pi_T) = \ell_T(\pi_T)$ and where the distribution of $Y_{k+1}$ given $(\pi_k, u_k)$ is given by \eqref{eq:Y_belief}.

The value functions $V_k^{\beta,g}$ and $V_k^{\beta,\ell}$ are, in general, not equal since the belief-state forms of the smoother entropy in \eqref{eq:secondBeliefAdditiveForm} and \eqref{eq:firstBeliefAdditiveForm} used to construct \eqref{eq:secondActiveEstimationMDP} and \eqref{eq:firstActiveEstimationMDP} breakdown the smoother entropy into different increments.
Indeed, the explicit separation of the conditional entropy $H(X_0|Y_0)$ in \eqref{eq:firstBeliefAdditiveForm} results in the following corollary to Theorem \ref{theorem:activeEstimationMDP} describing the relationship between $V_k^{\beta,g}$ and $V_k^{\beta,\ell}$ at $k = 0$.
\begin{corollary}
\label{corollary:valueFunctions}
Consider \eqref{eq:activeEstimation} with any $\beta \in (-\infty, \infty)$.
Then, $V_0^{\beta,g}(\pi_0) = V_0^{\beta,\ell}(\pi_0) + \beta H(X_0|Y_0)$.
\end{corollary}
\begin{IEEEproof}
    Note that $V_0^{\beta,g}$ is equal to the infimum in \eqref{eq:secondActiveEstimationMDP} (under the constraints \eqref{eq:beliefStateConstraints}), whilst $V_0^{\beta,\ell}$ is equal to the infimum in \eqref{eq:firstActiveEstimationMDP} (under the constraints \eqref{eq:beliefStateConstraints}). 
    Since Theorem \ref{theorem:activeEstimationMDP} gives that \eqref{eq:secondActiveEstimationMDP} and \eqref{eq:firstActiveEstimationMDP} are equal to \eqref{eq:activeEstimation}, we have that
    \begin{align*}
        V_0^{\beta,g}(\pi_0)
        &= \inf_{\mu \in \mathcal{P}} \left\{ \beta H_\mu(X^T | Y^T, U^{T-1}) + J(\mu) \right\}\\
        &= V_0^{\beta,\ell}(\pi_0) + \beta H(X_0|Y_0)
    \end{align*}
    subject to the constraints in \eqref{eq:activeEstimation}.
    The proof is complete.
\end{IEEEproof}

Regardless of any differences between the value functions $V_k^{\beta,g}$ and $V_k^{\beta,\ell}$, the next corollary to Theorem \ref{theorem:activeEstimationMDP} highlights that they must both describe belief-state policies solving \eqref{eq:activeEstimation}.

\begin{corollary}
\label{corollary:policy}
Consider \eqref{eq:activeEstimation} with any $\beta \in (-\infty, \infty)$.
If a policy $\bar{\mu}^{\beta*} =\{\bar{\mu}_k^{\beta*} : 0 \leq k < T\}$ satisfies
\begin{align*}
    \begin{split}
    \bar{\mu}_k^{\beta*}(\pi_k) 
    &= u_k^{\beta*}
    \in \arginf_{u_k \in \mathcal{U}} \{ \ell_k^\beta ( \pi_{k}, u_{k} ) \\
    &\; \qquad\qquad + E_{Y_{k+1}} [ V_{k+1}^{\beta,\ell}(\Pi(\pi_{k}, u_k, y_{k+1})) | \pi_k, u_k ] \}
    \end{split}
\end{align*}
for $0 \leq k < T$, or 
\begin{align*}
    \begin{split}
    \bar{\mu}_k^{\beta*}(\pi_k) 
    &= u_k^{\beta*}
    \in \arginf_{u_k \in \mathcal{U}} \{ g_k^\beta ( \pi_{k}, u_{k} ) \\
    &\; \qquad \qquad + E_{Y_{k+1}} [ V_{k+1}^{\beta,g}(\Pi(\pi_{k}, u_k, y_{k+1})) | \pi_k, u_k ] \}
    \end{split}
\end{align*}
for $0 \leq k < T$, then it solves \eqref{eq:activeEstimation}.
\end{corollary}
\begin{IEEEproof}
    From Theorem \ref{theorem:activeEstimationMDP}, \eqref{eq:secondActiveEstimationMDP} and \eqref{eq:firstActiveEstimationMDP} are equivalent reformulations of \eqref{eq:activeEstimation} (with \eqref{eq:firstActiveEstimationMDP} being equivalent up to the additive constant $\beta H(X_0 | Y_0)$, which does not affect the optimising policy).
    Thus, policies solving \eqref{eq:secondActiveEstimationMDP} and \eqref{eq:firstActiveEstimationMDP}, and hence satisfying their associated dynamic programming equations \eqref{eq:secondEstValueFunction} and \eqref{eq:firstEstValueFunction}, also solve \eqref{eq:activeEstimation}. The proof is complete.
\end{IEEEproof}

Corollary \ref{corollary:policy} reinforces the key conclusion of Theorem \ref{theorem:activeEstimationMDP}, namely, that we can find optimal policies solving \eqref{eq:activeEstimation} by instead solving either \eqref{eq:secondActiveEstimationMDP} or \eqref{eq:firstActiveEstimationMDP} via (belief-state) dynamic programming.
Whilst solving dynamic programming equations for optimal policies is typically difficult, (approximate) solutions can be found when the underlying cost and value functions have the same structural properties as standard POMDPs of the form in \eqref{eq:standardPOMDP}.
Specifically, if either \eqref{eq:secondActiveEstimationMDP} or \eqref{eq:firstActiveEstimationMDP} have cost and value functions that are concave in the belief state, then we can employ standard POMDP techniques to solve \eqref{eq:activeEstimation} (cf.\ \cite{Araya2010} and \cite[Chapter 8]{Krishnamurthy2016}).
We therefore now investigate the structural properties of \eqref{eq:secondActiveEstimationMDP} and \eqref{eq:firstActiveEstimationMDP}.
%We will later use these results to find (approximate) solutions to our active estimation or obfuscation problem \eqref{eq:activeEstimation}.

\subsection{Structural Results}
\label{subsect:structResults}

We first examine the structure of the instantaneous and terminal cost functions of the first belief-state MDP \eqref{eq:secondActiveEstimationMDP}.
\begin{lemma}
\label{lemma:estConcave}
Consider \eqref{eq:secondActiveEstimationMDP} with any $\beta \in (-\infty, \infty)$.
The terminal cost function $g_T^\beta(\pi)$, and the instantaneous cost functions $g_k^\beta(\pi, u_k)$ for any $u_k \in \mathcal{U}$ and $0 \leq k < T$, are:
\begin{enumerate}[label = \roman*)]
    \item 
    Concave and continuous in $\pi \in \Delta$ for $\beta > 0$;
    \item
    Convex and continuous in $\pi \in \Delta$ for $\beta < 0$; and,
    \item
    Linear and continuous in $\pi \in \Delta$ for $\beta = 0$.
\end{enumerate}
\end{lemma}
\begin{IEEEproof}
For any $\beta \in (-\infty, \infty)$, the definition of $g_T^\beta$ gives
\begin{align*}
    g_T^\beta(\pi_{T})
    &= E_{X_{T}} \left[\beta\tilde{g}_T(\pi_T) + c_T (x_{T}) | \pi_{T} \right]\\
    %&= \tilde{g}_T(\pi_T) + E_{X_{T}} \left[c_T (x_{T}) | \pi_{T} \right]\\
    &= \beta H(X_T | y^T, u^{T-1}) + \sum_{x \in \mathcal{X}} \pi_T(x) c_T(x).
\end{align*}
The second term on the right-hand side of this equation is linear and continuous in $\pi_T$ for any $\beta \in (-\infty,\infty)$ (hence $g_T^\beta$ is linear when $\beta = 0$).
Since the sum of a concave (or convex) function with a linear function remains concave (resp. convex), the structure of $g_T^\beta$ when $\beta \neq 0$ is determined solely by the first term, i.e.\ $\beta H(X_T | y^T, u^{T-1})$.
Noting that $H(X_T | y^T, u^{T-1})$ is the entropy of the belief state $\pi_T$, which is concave and continuous in $\pi_T$ via standard results (cf.~\cite[Theorem 2.7.3]{Cover2006}), it follows that $\beta H(X_T | y^T, u^{T-1})$ is concave for $\beta > 0$ and convex for $\beta < 0$.
The lemma assertion for $g_T^\beta$ follows.

Similarly, for any $u_k \in \mathcal{U}$, $0 \leq k < T$, and $\beta \in (-\infty, \infty)$, the definition of $g_k^\beta$ gives that
\begin{align*}
    g_k^\beta(\pi_{k}, u_{k})
    &= E_{X_{k}} \left[\beta \tilde{g}(\pi_k, u_k) + c_k (x_{k}, u_{k}) | \pi_{k}, u_{k} \right]\\
    %&= \tilde{g}(\pi_k, u_k)  + E_{X_{k}} \left[c_k (x_{k}, u_{k}) | \pi_{k}, u_{k} \right]\\
    &= \beta H(X_k | X_{k+1}, y^k, u^k) + \sum_{x \in \mathcal{X}} \pi_k(x) c_k(x, u_k).
\end{align*}
The second term on the right-hand side is linear in $\pi_k$.
Hence, $g_k^\beta$ is linear when $\beta = 0$, and the structure of $g_k^\beta(\pi_{k}, u_{k})$ when $\beta \neq 0$ is determined by the first term, i.e.\ $\beta H(X_k | X_{k+1}, y^k, u^k)$.
Note $H(X_k | X_{k+1}, y^k, u^k)$ is a conditional entropy so it is continuous and concave in the (joint) pmf $p(x_k, x_{k+1} | y^k, u^k)$ (cf.\ \cite[Appendix A]{Globerson2007} or \cite[Facts 1.4.6 and 1.7.9]{Downarowicz2011}).
The pmf $p(x_k, x_{k+1} | y^k, u^k)$ is the joint predicted belief $\bar{\pi}_{k+1 | k}$, which is a linear function of $\pi_k$ for any $u_k \in \mathcal{U}$, as shown in \eqref{eq:bayesianPred}.
Thus, $H(X_k | X_{k+1}, y^k, u^k)$ is a concave function of a linear function of $\pi_k$, and so it is concave and continuous in $\pi_k$.
It follows that $\beta H(X_k | X_{k+1}, y^k, u^k)$ (and $g_k^\beta(\pi_{k}, u_{k})$) is concave in $\pi_k$ for $\beta > 0$, and convex in $\pi_k$ for $\beta < 0$.
The proof is complete.
\end{IEEEproof}

We next examine the structure of the instantaneous and terminal cost functions of the second belief-state MDP \eqref{eq:firstActiveEstimationMDP}.

\begin{lemma}
\label{lemma:estConvex}
Consider \eqref{eq:firstActiveEstimationMDP} with any $\beta \in (-\infty, \infty)$.
The terminal cost function $\ell_T(\pi)$ is linear in $\pi$, and the instantaneous cost functions $\ell_k^\beta(\pi, u_k)$ for any $u_k \in \mathcal{U}$ and $0 \leq k < T$ are:
\begin{enumerate}[label = \roman*)]
    \item
    Convex and continuous in $\pi \in \Delta$ for $\beta > 0$;
    \item
    Concave and continuous in $\pi \in \Delta$ for $\beta < 0$; and,
    \item
    Linear and continuous in $\pi \in \Delta$ for $\beta = 0$.
\end{enumerate}
\end{lemma}
\begin{IEEEproof}
The definition of $\ell_T$ implies that
\begin{align*}
    \ell_T(\pi_{T}) = E_{X_{T}} \left[c_T (x_{T}) | \pi_{T} \right] = \sum_{x \in \mathcal{X}} \pi_T(x) c_T(x),
\end{align*}
which is linear and continuous in $\pi_T$, regardless of $\beta$.

The definition of $\ell_k^\beta$ for any $\beta \in (-\infty,\infty)$ gives that
\begin{align}\notag
    &\ell_k^\beta(\pi_{k}, u_{k})\\\notag
%    &= E_{Y_{k+1},X_{k}} \left[\left. \tilde{\ell}(\pi_k, u_k,y_{k+1}) + c_k (x_{k}, u_{k}) \right| \pi_{k}, u_{k} \right]\\\notag
    &= E_{Y_{k+1}} \left[\left. \beta \tilde{\ell}_1(\pi_k, u_k, y_{k+1}) \right| \pi_k, u_k \right] \\\notag
    &\quad- \beta \tilde{\ell}_2(\pi_k, u_k) + \beta \tilde{\ell}_3(\pi_k, u_k) + E_{X_k} \left[ \left. c_k (x_{k}, u_{k}) \right| \pi_{k}, u_{k} \right] \\\notag
    &= \beta H(X_{k+1} | Y_{k+1}, y^k, u^k) - \beta H(X_{k+1}|y^k, u^k) \\\notag
    &\quad+ \beta H(X_{k+1}|X_k, y^k, u^k) + E_{X_k} \left[ \left. c_k (x_{k}, u_{k}) \right| \pi_{k}, u_{k} \right]\\\label{eq:concave_proof_step1}
    \begin{split}
    &= \beta H(X_{k+1}|X_k, y^k, u^k) - \beta I(X_{k+1}; Y_{k+1} | y^{k}, u^k) \\
    &\quad+ \sum_{x \in \mathcal{X}} \pi_k(x) c_k(x, u_k)
    \end{split}
\end{align}
where the last equality holds since $I(X_{k+1}; Y_{k+1} | y^{k}, u^{k}) = H(X_{k+1}|y^k, u^k) - H(X_{k+1}|Y_{k+1}, y^k, u^k)$.
For any $u_k \in \mathcal{U}$, the first and third terms in \eqref{eq:concave_proof_step1} are linear and continuous in $\pi_k$ for any $\beta \in (-\infty, \infty)$, as shown in \eqref{eq:first_cond_cost_entropy} for the first term.
Hence, $\ell_k^\beta$ is linear when $\beta = 0$.
Furthermore, since the sum of a concave (or convex) function with linear functions remains concave (resp. convex), the structure of $\ell_k^\beta(\pi_{k}, u_{k})$ when $\beta \neq 0$ is thus determined solely by the second term in \eqref{eq:concave_proof_step1}, i.e.\ $-\beta I(X_{k+1}; Y_{k+1} | y^{k}, u^{k})$.

For $u_k \in \mathcal{U}$, $-I(X_{k+1}; Y_{k+1} | y^{k}, u^{k})$ is convex in $\pi_k$ since:
\begin{enumerate}
    \item
    $-I(X_{k+1}; Y_{k+1} | y^{k}, u^{k})$ is convex and continuous in $p(x_{k+1} | y^{k}, u^{k})$ via \cite[Theorem 2.7.4]{Cover2006} with the conditional pmf $p(y_{k+1} | x_{k+1}, y^{k}, u^{k}) = p(y_{k+1} | x_{k+1}, u_k)$ fixed and determined by the measurement kernel \eqref{eq:obsProcess}; and,
    \item
    $p(x_{k+1} | y^{k}, u^{k})$ is a linear function of $\pi_k$ since it is the marginal of the joint predicted belief $\bar{\pi}_{k+1|k}$ from \eqref{eq:bayesianPred}.
\end{enumerate}
That is, $-I(X_{k+1}; Y_{k+1} | y^{k}, u^{k})$ is convex in a linear function of $\pi_k$, and thus is convex and continuous in $\pi_k$.
Hence, $-\beta I(X_{k+1}; Y_{k+1} | y^{k}, u^{k})$ (and $\ell_k^\beta$) is convex in $\pi_k$ when $\beta > 0$, and concave when $\beta < 0$, completing the proof.
\end{IEEEproof}

Lemmas \ref{lemma:estConcave} and \ref{lemma:estConvex} are surprising because they show that the terminal and instantaneous cost functions of the two belief-state MDPs \eqref{eq:secondActiveEstimationMDP} and \eqref{eq:firstActiveEstimationMDP} have different structural properties, despite both being reformulations of \eqref{eq:activeEstimation}.
Specifically, the terminal and instantaneous costs of \eqref{eq:secondActiveEstimationMDP} are concave (convex) when $\beta > 0$ (resp. $\beta < 0$), whilst the terminal and instantaneous costs of \eqref{eq:firstActiveEstimationMDP} are convex (concave) when $\beta > 0$ (resp. $\beta < 0$).
Since standard POMDP solution techniques require the terminal and instantaneous costs of belief-state MDP reformulations to be concave in the belief state (cf.\ \cite{Araya2010,Krishnamurthy2016}), the convexity results of Lemmas \ref{lemma:estConcave} and \ref{lemma:estConvex} do not assist us in solving \eqref{eq:activeEstimation}.
However, the concavity results of Lemmas \ref{lemma:estConcave} and \ref{lemma:estConvex} lead directly to the following theorem establishing that the value function $V_k^{\beta,g}$ of \eqref{eq:secondActiveEstimationMDP} is concave for $\beta \geq 0$, whilst the value function $V_k^{\beta,\ell}$ of \eqref{eq:firstActiveEstimationMDP} is concave for $\beta \leq 0$.

%Since standard POMDP techniques require the terminal and instantaneous costs of the belief-state MDP reformulations to be concave in the belief state (cf.\ \cite{Araya2010,Krishnamurthy2016}), the convexity results do not further assist us in solving \eqref{eq:activeEstimation}.
%However, the concavity results suggest that a belief-state MDP reformulation of our active obfuscation problem \eqref{eq:activeObfuscation} using the belief-state form of the smoother entropy in \eqref{eq:firstBeliefAdditiveForm} may have useful concavity properties since it involves maximising, rather than minimising, the smoother entropy.

%The concavity of the instantaneous costs $g_k^e$ established in Lemma \ref{lemma:estConcave} is nontrivial because conditional entropies are, in general, only concave in the joint distribution of their arguments (cf.\ \cite[Appendix A]{Globerson2007}) whilst here we consider the concavity of $H(X_k | X_{k+1}, y^k, u^k)$ in the marginal distribution of $X_k$ (i.e., the belief-state).
\begin{theorem}
\label{theorem:estConcave}
 Consider \eqref{eq:activeEstimation} for any $\beta \in (-\infty, \infty)$.
 \begin{enumerate}[label = \roman*)]
     \item
     If $\beta \geq 0$, then the value function $V_k^{\beta,g}(\pi_k)$ of the first belief-state MDP reformulation of \eqref{eq:activeEstimation} in \eqref{eq:secondActiveEstimationMDP} is concave in $\pi_k \in \Delta$ for all $0 \leq k \leq T$.
     \item
     Conversely, if $\beta \leq 0$, then the value function $V_k^{\beta,\ell}(\pi_k)$ of the second belief-state MDP reformulation of \eqref{eq:activeEstimation} in \eqref{eq:firstActiveEstimationMDP} is concave in $\pi_k \in \Delta$ for all $0 \leq k \leq T$.
 \end{enumerate}
\end{theorem}
\begin{IEEEproof}
The assertions follow from \cite[Theorem 8.4.1]{Krishnamurthy2016} due to the concavity and continuity of $g_k^\beta$ and $g_T^\beta$ established in Lemma \ref{lemma:estConcave} for $\beta \geq 0$, and the concavity and continuity of $\ell_k^\beta$ and $\ell_T$ established in Lemma \ref{lemma:estConvex} for $\beta \leq 0$.
\end{IEEEproof}

The structural results of Lemmas \ref{lemma:estConcave} and \ref{lemma:estConvex}, and Theorem \ref{theorem:estConcave} are surprising because they imply that \eqref{eq:activeEstimation} can be reformulated as a belief-state MDP with the same concavity properties as standard POMDPs of the form in \eqref{eq:standardPOMDP}, regardless of whether we are minimising or maximising the smoother entropy via $\beta \geq 0$ or $\beta \leq 0$, respectively.
Specifically, if we wish to minimise the smoother entropy via $\beta \geq 0$, the belief-state MDP reformulation  \eqref{eq:secondActiveEstimationMDP} has the same concavity properties as standard POMDPs of the form in \eqref{eq:standardPOMDP}.
Conversely, if we wish to maximise the smoother entropy via $\beta \leq 0$, the alternative belief-state MDP reformulation  \eqref{eq:firstActiveEstimationMDP} has the same concavity properties as standard POMDPs of the form in \eqref{eq:standardPOMDP}.
We note, however, that despite Lemmas \ref{lemma:estConcave} and \ref{lemma:estConvex} establishing that the cost functions $g_k^\beta$ for $\beta < 0$ and $\ell_k^\beta$ for $\beta > 0$ are convex in the belief state, Theorem \ref{theorem:estConcave} does not characterise the structure of the corresponding value functions $V_k^{\beta,g}$ for $\beta < 0$ or $V_k^{\beta,\ell}$ for $\beta > 0$.
Such a characterisation will prove unnecessary since the structural results we have established are already sufficient to enable the solution of \eqref{eq:activeEstimation} using standard POMDP techniques regardless of whether $\beta \geq 0$ or $\beta \leq 0$.
%Furthermore, such a characterisation would be of little practical value because as noted in \cite{Fehr2018}, few (approximate) solution techniques exist for POMDPs with nonconcave cost functions.

\subsection{Apparent Paradox of Convex Instantaneous Cost Functions}

%(which enables the initial entropy $\beta H(X_0|Y_0)$ to be omitted from its value function).
%Similarly, Proposition \ref{proposition:estConvex} considers the MDP reformulation of active estimation based on the belief-state form of the smoother entropy in \eqref{eq:firstBeliefAdditiveForm} whilst Proposition \ref{proposition:obfConvex} considers the MDP reformulation of active obfuscation based on the belief-state form of the smoother entropy in \eqref{eq:secondBeliefAdditiveForm}.
The convexity (concavity) of the instantaneous cost functions $\ell_k^\beta$ in the belief state for $\beta > 0$ (resp. $\beta < 0$) established in Lemma \ref{lemma:estConvex} is particularly surprising because the minimisation (resp.\ maximisation) of most standard state-uncertainty measures leads to concave (resp.\ convex) instantaneous cost functions (cf.\ \cite[Section 8.4.3]{Krishnamurthy2016} and \cite{Fehr2018}).
The concavity (convexity) of standard belief-state cost functions associated with minimising (maximising) standard state-uncertainty measures reflects the intuition that the belief states that represent the least (most) state uncertainty correspond to the vertices (resp.\ centre) of the probability simplex $\Delta$.
Upon first inspection, our structural results for the instantaneous cost functions $\ell_k^\beta$ might appear to contradict this intuition.
However, there is no contradiction because the instantaneous cost functions $\ell_k^\beta$ are not themselves directly interpretable as measures of state uncertainty --- they only correspond to the smoother entropy $H_\mu(X^T | Y^T, U^{T-1})$ after taking expectations and when combined with the initial constant $H(X_0 | Y_0)$ (which can separated and omitted during the optimisation in \eqref{eq:firstActiveEstimationMDP}).
%Indeed, as shown in the proof of Lemma \ref{lemma:estConvex} (cf.\ \eqref{eq:concave_proof_step1}), $\ell_k^\beta$ are the weighted sum of both an uncertainty term $H(X_{k+1} | X_k, y^k, u^k)$ that is linear in $\pi_k$ (and hence obeys the standard intuition due to being both concave and convex) and a (negative) information term $-I(X_{k+1}; Y_{k+1} | y^k, u^k)$.
%The information term $-I(X_{k+1}; Y_{k+1} | y^k, u^k)$ is not actually 

For example, consider \eqref{eq:activeEstimation} in the simple case where $T = 1$, $\beta = 1$, and $c_T$ and $c_k$ are zero functions, then \eqref{eq:firstActiveEstimationMDP} (and its proof in Theorem \ref{theorem:activeEstimationMDP} using \eqref{eq:firstBeliefAdditiveForm}) implies that
\begin{align*}
    \inf_{\mu \in \mathcal{P}} H_\mu(X^1 | Y^1, U_0)
    %&=\; \inf_{\bar{\mu}} E_{\bar{\mu}} \left[ g_1(\pi_1) + g_0^1 \left( \pi_0, u_0 \right) \right]\\
    %&=\; \inf_{\bar{\mu}} E_{\bar{\mu}} \left[ H (X_1 | y^1, u_0) + H (X_0 | X_1, y_0, u_0) \right]\\
    &= H(X_0 | Y_0) + \inf_{\mu \in \mathcal{P}} E_\mu \left[ \ell_0^1 \left( \pi_0, u_0 \right) \right]\\
    &= \inf_{\mu \in \mathcal{P}} E_\mu \left[ H(X_0 | y_0) + \ell_0^1 \left( \pi_0, u_0 \right) \right]
    %&\;= H(X_0 | Y_0) + \inf_{\bar{\mu}} E_{\bar{\mu}} \left[ H (X_1 | Y_1, y_0, u_0) - H (X_1 | y_0, u_0) + H (X_1 | X_0, y_0, u_0) \right]\\
    %&= H(X_0 | Y_0) + \inf_\mu E_\mu \left[ H (X_1 | X_0, y_0, u_0) - I (X_1 ; Y_1 | y_0, u_0) \right]\\
\end{align*}
subject to the constraints in \eqref{eq:activeEstimation} with the second line holding because $H(X_0 | y_0) = - \sum_{x\in \mathcal{X}} \pi_0(x) \log \pi_0(x)$ is the entropy of $\pi_0$.
From the last expression, we see that whilst $\ell_0^1 \left( \pi_0, u_0 \right)$ is convex in $\pi_0$ for any $u_0 \in \mathcal{U}$ via Lemma \ref{lemma:estConvex}, it is still possible for the total cost within the expectation, i.e.~$H(X_0 | y_0) + \ell_0^1 \left( \pi_0, u_0 \right)$, to be concave in $\pi_0$ due to the concavity of the entropy $H(X_0 | y_0)$ in $\pi_0$ \cite[Theorem 2.7.3]{Cover2006}.

\subsection{Solving \eqref{eq:activeEstimation} via Standard POMDP Techniques}
\label{sec:approx}

The results developed in Lemmas \ref{lemma:estConcave} and \ref{lemma:estConvex}, and Theorem \ref{theorem:estConcave} are practically significant because they enable the solution of \eqref{eq:activeEstimation} using standard POMDP techniques, regardless of whether $\beta \geq 0$ or $\beta \leq 0$.
%Furthermore, use of existing POMDP algorithms to solve \eqref{eq:activeEstimation} requires that its cost and value functions are piecewise-linear concave (PWLC) in the belief state (cf.\ \cite{Araya2010} and \cite[Chapter 8.4.4]{Krishnamurthy2016}).
Here, we present one such standard technique that was originally introduced in \cite{Araya2010} for infinite-horizon discounted $\rho-$POMDPs, but which we shall show also yields tractable bounded-error approximate solutions to our finite-horizon undiscounted problem \eqref{eq:activeEstimation} by exploiting \eqref{eq:secondActiveEstimationMDP} for $\beta \geq 0$ and \eqref{eq:firstActiveEstimationMDP} for $\beta \leq 0$.
This approach involves:
\begin{enumerate}
    \item 
     Constructing bounded-error piecewise-linear concave (PWLC) approximations of the concave costs $g_k^\beta$ for $\beta > 0$ and $\ell_k^\beta$ for $\beta < 0$; and,
     \item
     Using the PWLC approximations of $g_k^\beta$ and $\ell_k^\beta$ with standard POMDP algorithms to solve \eqref{eq:secondEstValueFunction} or \eqref{eq:firstEstValueFunction} for PWLC approximations of the value functions $V_k^{\beta,g}$ for $\beta > 0$ and $V_k^{\beta, \ell}$ for $\beta < 0$.
\end{enumerate}
To present this approach, we shall assume that the measurement space $\mathcal{Y}$ is finite (e.g., as given or obtained by discretising a continuous space).

\subsubsection{Bounded-Error PWLC Cost Approximations}

The concavity of the cost functions $g_k^\beta$ and $\ell_k^\beta$ established in Lemmas \ref{lemma:estConcave} and \ref{lemma:estConvex} allows us to approximate them using PWLC functions.
Specifically, let us consider a finite set $\Xi \subset \Delta$ of \emph{base points} $\xi \in \Xi$ at which the gradients $\nabla_\pi g_k^\beta(\xi, u)$ and $\nabla_\pi \ell_k^\beta(\xi, u)$ of $g_k^\beta(\cdot, u)$ and $\ell_k^\beta(\cdot, u)$, respectively, are well defined for all $u \in \mathcal{U}$.
For each control $u \in \mathcal{U}$, the tangent hyperplane to $g_k^\beta(\cdot, u)$ at $\xi \in \Xi$ is
\begin{align*}
    \omega_{k,\xi}^{g,u} (\pi)
    \triangleq g_k^\beta(\xi, u) + \left< (\pi - \xi), \nabla_\pi g_k^\beta(\xi, u) \right>
    = \left< \pi, \alpha_{k,\xi}^{g,u} \right>
\end{align*}
and the tangent hyperplane to $\ell_k^\beta(\cdot, u)$ at $\xi \in \Xi$ is
\begin{align*}
    \omega_{k,\xi}^{\ell,u} (\pi)
    \triangleq \ell_k^\beta(\xi, u) + \left< (\pi - \xi), \nabla_\pi \ell_k^\beta(\xi, u) \right>
    = \left< \pi, \alpha_{k,\xi}^{\ell,u} \right>
\end{align*}
for $\pi \in \Delta$ where $\left< \cdot, \cdot \right>$ denotes the inner product, and $\alpha_{k,\xi}^{g,u} \triangleq g_k^\beta(\xi, u) + \nabla_\pi g_k^\beta(\xi, u) - \left< \xi, \nabla_\pi g_k^\beta(\xi, u) \right> \in \mathbb{R}^{N_x}$ and $\alpha_{k,\xi}^{\ell,u} \triangleq \ell_k^\beta(\xi, u) + \nabla_\pi \ell_k^\beta(\xi, u) - \left< \xi, \nabla_\pi \ell_k^\beta(\xi, u) \right> \in \mathbb{R}^{N_x}$ (with the addition of a vector and a scalar here meaning the addition of the scalar to all elements of the vector).
Due the the concavity results in Lemmas \ref{lemma:estConcave} and \ref{lemma:estConvex}, the hyperplanes $\omega_{k,\xi}^{g,u}$ and $\omega_{k,\xi}^{\ell,u}$ form (upper bound) PWLC approximations $\hat{g}_k^\beta$ and $\hat{\ell}_k^\beta$ to $g_k^\beta$ and $\ell_k^\beta$ for $\beta > 0$ and $\beta < 0$, respectively. 
That is,
\begin{align*}
    \hat{g}_k^\beta(\pi, u)
    \triangleq \min_{\xi \in \Xi} \left< \pi, \alpha_{k,\xi}^{g,u} \right> 
    \geq g_k^\beta(\pi, u)
\end{align*}
for $\beta > 0$, and
\begin{align*}
    \hat{\ell}_k^\beta(\pi, u) \triangleq \min_{\xi \in \Xi} \left< \pi, \alpha_{k,\xi}^{\ell,u} \right> \geq \ell_k^\beta(\pi, u)
\end{align*}
for $\beta < 0$.
PWLC approximations of the concave terminal costs $g_T^\beta$ for $\beta > 0$ are constructed in an identical manner without the need to consider the controls (since $\ell_T$ is linear, no approximations are needed).
As shown in the following lemma, the approximation errors associated with $\hat{g}_k^\beta$ and $\hat{\ell}_k^\beta$ are bounded for $\beta > 0$ and $\beta < 0$, respectively.

\begin{lemma}
    \label{lemma:bounds}
    Consider the set of base points $\Xi$ and associated PWLC approximations $\hat{g}_k^\beta$ and $\hat{g}_T^\beta$ for $\beta \geq 0$, and $\hat{\ell}_k^\beta$ for $\beta \leq 0$ for all $0 \leq k < T$.
    Then there exists scalar constants $\kappa^g, \kappa^\ell > 0$, and $\eta^g, \eta^\ell \in (0,1)$ such that the errors in the approximations $\hat{g}_k^\beta$ and $\hat{\ell}_k^\beta$ are bounded for $\beta > 0$ and $\beta < 0$, respectively; that is, 
    $
        |g_k^\beta(\pi,u) - \hat{g}_k^\beta(\pi,u)| \leq \kappa^g(\delta_\Xi)^{\eta^g}
    $
    for $\beta > 0$, and
    $
        |\ell_k^\beta(\pi,u) - \hat{\ell}_k^\beta(\pi,u)| \leq \kappa^\ell(\delta_\Xi)^{\eta^\ell}
    $
    for all $0 \leq k \leq T$, all $\pi \in \Delta$, and all $u \in \mathcal{U}$ where $\delta_\Xi \triangleq \min_{\pi \in \Delta} \max_{\xi \in \Xi} \| \pi - \xi \|_1$ is the sparsity of the base-point set $\Xi$ and $\|\cdot\|_1$ denotes the $l^1$-norm.
\end{lemma}
\begin{IEEEproof}
    Recall that a function $f : \mathcal{D} \rightarrow \mathbb{R}$ with $\mathcal{D} \subset \mathbb{R}^{N_x}$ is $\eta$-H{\"o}lder continuous on $\mathcal{D}$ if these exists constants $\eta \in (0,1]$ and $K_\eta >0$ such that $|f(x) - f(y)| \leq K_\eta \|x - y\|_1^\eta$ for all $x,y \in \mathcal{D}$ \cite{Araya2010}.
    We note that the entropy function $f(\pi) = -\sum_{x \in \mathcal{X}} \pi(x) \log \pi(x)$ is $\eta$-H{\"o}lder continuous on $\Delta$ with $\eta < 1$ and the convention $0 \log 0 = 0$ (cf.\ \cite[Example 1.1.4]{Fiorenza2017} and \cite[p. 7]{Araya2010}).
    Furthermore, continuous linear functions are $\eta$-H{\"o}lder continuous, as are the sums, differences, and compositions of $\eta$-H{\"o}lder continuous functions (cf.~\cite[Propositions 1.2.1 and 1.2.2]{Fiorenza2017}).
    Thus, for each control $u_k \in \mathcal{U}$, the functions $g_k^\beta$ and $\ell_k^\beta$ are $\eta$-H{\"o}lder continuous in $\pi_k$ since each term in $g_k^\beta$ and $\ell_k^\beta$ is either linear in $\pi_k$, or can be expressed as the composition of a linear function and the entropy function (e.g.\  via \eqref{eq:bayesianPred}).
    The $\eta$-H{\"o}lder continuity of $g_k^\beta$ and $\ell_k^\beta$ combined with their continuity and concavity properties established in Lemmas \ref{lemma:estConcave} and \ref{lemma:estConvex} for $\beta > 0$ and $\beta < 0$, respectively, imply that $g_k^\beta$ and $\ell_k^\beta$ satisfy the conditions of \cite[Theorem 4.3]{Araya2010} for each control $u_k \in \mathcal{U}$, and for $\beta > 0$ and $\beta < 0$, respectively.
    The lemma assertion follows from \cite[Theorem 4.3]{Araya2010} (noting that here we equivalently consider upper bounds on concave functions rather than lower bounds on convex functions).
\end{IEEEproof}

\subsubsection{PWLC Dynamic Programming and Error Bounds}

Standard POMDP algorithms provide a means of solving belief-state dynamic programming equations when the cost and value functions involved are PWLC in the belief state (cf.\ \cite{Araya2010} and \cite[Chapter 8.4.4]{Krishnamurthy2016}).
Hence, by replacing the costs $g_k^\beta$ and $\ell_k^\beta$ in the dynamic programming equations of \eqref{eq:secondEstValueFunction} and \eqref{eq:firstEstValueFunction} with the PWLC approximations $\hat{g}_k^\beta$ and $\hat{\ell}_k^\beta$ for $\beta > 0$ and $\beta < 0$, respectively, the equations can be solved for approximate value functions $\hat{V}_k^{\beta,g}$ and $\hat{V}_k^{\beta,\ell}$ using standard POMDP algorithms.
Under the assumption that $\mathcal{Y}$ is finite, the resulting approximate value functions are PWLC, which standard POMDP algorithms can exploit by operating directly on the sets of vectors $\{\alpha_{k,\xi}^{g,u} : \xi \in \Xi,\, u \in \mathcal{U}\}$ and $\{\alpha_{k,\xi}^{\ell,u} : \xi \in \Xi,\, u \in \mathcal{U}\}$ that define $\hat{g}_k^\beta$ and $\hat{\ell}_k^\beta$ (see \cite[Chapter 7.5]{Krishnamurthy2016} and \cite[Section 3.3]{Araya2010} for details of these algorithms and their inherent requirement for concavity of the cost and value functions in the belief state).
The following proposition shows that the resulting value function errors are bounded.

\begin{proposition}
    \label{proposition:bounds}
    Consider the set of base points $\Xi$, the PWLC approximations $\hat{g}_k^\beta$ for $\beta > 0$ and $\hat{\ell}_k^\beta$ for $\beta < 0$, and the associated approximate value functions $\hat{V}_k^{\beta,g}$ for $\beta > 0$ and $\hat{V}_k^{\beta,\ell}$ for $\beta < 0$.
    Then there exists scalar constants $\kappa^g, \kappa^\ell > 0$, and $\eta^g, \eta^\ell \in (0,1)$ such that
    \begin{align}
        \label{eq:actEstValueBound}
        \| V_k^{\beta,g} - \hat{V}_k^{\beta,g}\|_\infty
    \leq (T - k + 1)\kappa^g (\delta_\Xi)^{\eta^g}
    \end{align}
    for $\beta > 0$ and $0 \leq k \leq T$, and
    \begin{align}
        \label{eq:actObfValueBound}
        \| V_k^{\beta,\ell} - \hat{V}_k^{\beta,\ell}\|_\infty
    \leq (T - k + 1)\kappa^\ell (\delta_\Xi)^{\eta^\ell}
    \end{align}
    for $\beta < 0$ and $0 \leq k \leq T$, where $\|\cdot\|_\infty$ denotes the $L^\infty$-norm.
\end{proposition}
\begin{IEEEproof}
    Consider first any $\beta > 0$ and the PWLC approximations $\hat{g}_k^\beta$ and $\hat{V}_k^{\beta,g}$.
    We prove \eqref{eq:actEstValueBound} via (backwards) induction on $k$.
    For $k = T$, \eqref{eq:actEstValueBound} holds via Lemma \ref{lemma:bounds} since $V_T^{\beta,g} = g_T$ and $\hat{V}_T^{\beta,g} = \hat{g}_T$.
    Let $\mathcal{T}$ denote the dynamic programming mapping using $g_k^\beta$ in the sense that
    \begin{align*}
        (\mathcal{T}V_{k+1}^{\beta,g})(\pi_k) 
        &\triangleq \inf_{u_{k} \in \mathcal{U}} \{ g_k^\beta ( \pi_k, u_{k} ) \\
        &\qquad + E_{Y_{k+1}} [ V_{k+1}^{\beta,g}(\Pi(\pi_{k}, u_{k}, y_{k+1})) | \pi_k, u_{k} ] \},
    \end{align*}
    for $\pi_k \in \Delta$. Similarly, let $\hat{\mathcal{T}}$ denote the mapping with $\hat{g}_k^\beta$, i.e.,
    \begin{align*}
        (\hat{\mathcal{T}}V_{k+1}^{\beta,g})(\pi_k) 
        &\triangleq \inf_{u_{k} \in \mathcal{U}} \{ \hat{g}_k^\beta ( \pi_k, u_{k} ) \\
        &\qquad + E_{Y_{k+1}} [ V_{k+1}^{\beta,g}(\Pi(\pi_{k}, u_{k}, y_{k+1})) | \pi_k, u_{k} ] \}
    \end{align*}
    for $\pi_k \in \Delta$. 
    Then, assuming that \eqref{eq:actEstValueBound} holds for times $T-1, \ldots, k+1$, at time $k$ we have that
    \begin{align*}
         &\| V_k^{\beta,g} - \hat{V}_k^{\beta,g}\|_\infty\\
         &\quad=  \| \mathcal{T} V_{k+1}^{\beta,g} - \hat{\mathcal{T}}\hat{V}_{k+1}^{\beta,g}\|_\infty\\
         &\quad\leq \| \mathcal{T} \hat{V}_{k+1}^{\beta,g} - \hat{\mathcal{T}} \hat{V}_{k+1}^{\beta,g}\|_\infty + \| \mathcal{T} V_{k+1}^{\beta,g} - \mathcal{T} \hat{V}_{k+1}^{\beta,g}\|_\infty\\
         &\quad\leq \kappa^g (\delta_\Xi)^{\eta^g} + \| \mathcal{T} V_{k+1}^{\beta,g} - \mathcal{T} \hat{V}_{k+1}^{\beta,g}\|_\infty\\
         &\quad\leq \kappa^g (\delta_\Xi)^{\eta^g} + \| V_{k+1}^{\beta,g} -  \hat{V}_{k+1}^{\beta,g}\|_\infty\\
         &\quad\leq (T - k + 1) \kappa^g (\delta_\Xi)^{\eta^g}
    \end{align*}
    where the first equality holds by definition of $\mathcal{T}$ and $\hat{\mathcal{T}}$; the first inequality is the triangle inequality; the second inequality holds via Lemma \ref{lemma:bounds} since $\mathcal{T}$ and $\hat{\mathcal{T}}$ differ in their use of $g_k^\beta$ and $\hat{g}_k^\beta$; the third inequality holds due to the monotonicity and constant-shift properties of the dynamic programming operator (cf.~\cite[Lemmas 1.1.1 and 1.1.2]{Bertsekas2012} and the argument in the convergence/contraction proof of \cite[Proposition 1.2.6]{Bertsekas2012}); and, the last inequality follows from the induction hypothesis.
    The proof of \eqref{eq:actEstValueBound} via induction is complete.
    With \eqref{eq:actObfValueBound} proved using an identical argument for $\beta < 0$, the proof is complete.
\end{IEEEproof}

\subsubsection{Complexity and Extensions}
As noted in \cite{Araya2010}, the (time) complexity of finding approximate solutions to belief-state MDPs with concave cost functions using PWLC approximations is only greater than that of solving standard POMDPs due to the size (number of base points) of the PWLC approximations.
Thus, given a set of base points, the complexity of finding a PWLC approximate solution to \eqref{eq:activeEstimation} scales the same as that of solving a standard finite-horizon POMDP with respect to the horizon $T$, and the number of states, controls, and measurements (see \cite{Hauskrecht2000,Krishnamurthy2016} and references therein for results on the complexity of solving standard POMDPs).
In practice, however, there is a trade-off between complexity and approximation error since Lemma \ref{lemma:bounds} and Proposition \ref{proposition:bounds} imply that the error in PWLC approximations becomes (arbitrarily) small through the selection of (many) base points that decrease $\delta_\Xi$.
The problem of optimising this trade-off is largely open \cite{Araya2010}, but recent point-based solvers (e.g.\ \cite{Walraven2019} for finite-horizon POMDPs and \cite{Garg2019,Kurniawati2008} for infinite-horizon POMDPs) provide insight via reachability results.

Importantly, our novel reformulations and structural results in Theorems \ref{theorem:activeEstimationMDP} and \ref{theorem:estConcave}, and Lemmas \ref{lemma:estConcave} and \ref{lemma:estConvex}, ensure that future developments in techniques for solving standard POMDPs will enable the solution of \eqref{eq:activeEstimation} with growing accuracy and  increasingly complicated state, control, and measurement processes.
For example, since the reformulations of \eqref{eq:activeEstimation} as belief-state MDPs in Theorem \ref{theorem:activeEstimationMDP} hold for continuous measurement spaces $\mathcal{Y}$, they enable the solution of \eqref{eq:activeEstimation} with continuous measurements using recent techniques for solving standard POMDPs with continuous measurement spaces (such as those presented in \cite{Hoerger2021}).
Furthermore, since our novel reformulations and structural results of Theorems \ref{theorem:activeEstimationMDP} and \ref{theorem:estConcave}, and Lemmas \ref{lemma:estConcave} and \ref{lemma:estConvex} hold without any specialised assumptions, we expect similar results to hold for various extensions of \eqref{eq:activeEstimation} such as when the set of controls $\mathcal{U}$ is state dependent, or when the horizon $T$ is infinite (with appropriate discounting or averaging of $J(\mu)$ and the smoother entropy).

We next discuss operational interpretations of \eqref{eq:activeEstimation}.

\section{Operational Interpretations and Comparisons}
\label{sec:operational}

In this section, we consider operational interpretations of \eqref{eq:activeEstimation}, and compare the smoother entropy with other state-uncertainty measures given the results of Section \ref{sec:activeEst}.

\subsection{Active State Trajectory Estimation}

The first operational interpretation of \eqref{eq:activeEstimation} we consider is within the context of controlling a partially observed stochastic system to aid in the estimation of its state trajectory.
Such active estimation problems arise in controlled sensing and target tracking \cite{Krishnamurthy2016,Chattopadhyay2018,Zois2014,Zois2017,Kartik2018}, uncertainty-aware robot navigation \cite{Nardi2019}, robot exploration \cite{Thrun2005}, and active SLAM \cite{Valencia2012,Roy2005}.

\begin{figure}[t!]
    \centering
    \includegraphics[width=\columnwidth]{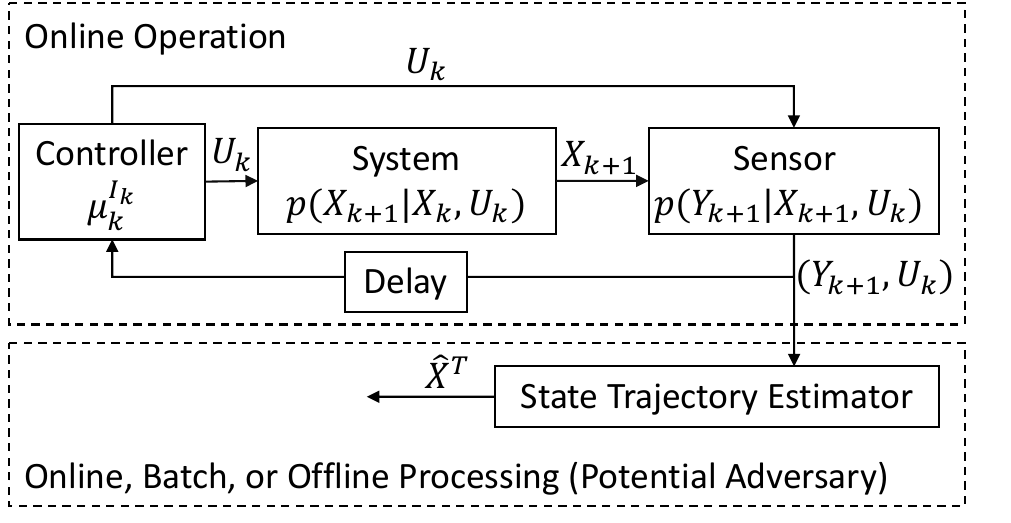}
    \caption{Active state trajectory estimation or obfuscation interpretation of \eqref{eq:activeEstimation}.}
    \label{fig:activeEstimation}
\end{figure}

Consider the setting shown in Fig.~\ref{fig:activeEstimation} in which a system and its attached sensors are controlled (online in real-time) via output feedback.
There is also a state trajectory estimator that processes the measurements and controls to compute estimates $\hat{X}^T \in \mathcal{X}^{T+1}$ of the system's entire state trajectory $X^T$ over some horizon $T$.
The state trajectory estimator is any function $f : \mathcal{Y}^{T+1} \times \mathcal{U}^{T} \rightarrow \mathcal{X}^{T+1}$ that maps the measurements and controls to a state trajectory, either online sequentially (such as in a recursive Bayesian filter \cite{Elliott1995, Krishnamurthy2016}), incrementally in batches (such as in incremental smoothing and mapping \cite{Kaess2008}), or wholly offline (such as in the Viterbi algorithm or an exact Bayesian smoother \cite{Krishnamurthy2016}).

Let the minimum probability of error achievable by the state trajectory estimator be
\begin{align*}
	\epsilon
	\triangleq \min_{\hat{X}^T \in \{ f : \mathcal{Y}^{T+1} \times \mathcal{U}^{T} \rightarrow \mathcal{X}^{T+1}\}} P(X^T \neq \hat{X}^T).
\end{align*}
This minimum error probability provides a fundamental bound on the performance of \emph{any} (potentially non-Bayesian) state trajectory estimator, and is achieved by maximum \emph{a posteriori} (MAP) state trajectory estimator such as the Viterbi algorithm \cite{Feder1994}.
Importantly, Theorem 1 of \cite{Feder1994} implies that the smoother entropy provides upper and lower bounds on $\epsilon$ in the sense that
\begin{align}
    \label{eq:smootherErrorBounds}
    \Phi^{-1}(H(X^T | Y^T, U^{T-1}))
    \leq \epsilon
    \leq \phi^{-1}(H(X^T | Y^T, U^{T-1}))
\end{align}
where $\Phi^{-1}$ and $\phi^{-1}$ are the inverse functions of strictly monotonically increasing continuous functions (defined in \cite[Eq.\ (9) and Eq.\ (14)]{Feder1994}), and so are also strictly monotonically increasing.
Thus, within the setting of Fig. \ref{fig:activeEstimation}, solving \eqref{eq:activeEstimation} with $\beta > 0$ has the operational interpretation of designing policies that improve the fundamental achievable performance of the state trajectory estimator (by reducing the smoother entropy and hence both upper and lower bounds on $\epsilon$).

\subsection{Active State Trajectory Obfuscation}

The second interpretation of \eqref{eq:activeEstimation} we consider is within the context of controlling a system to hinder the estimation of its state trajectory.
Such active obfuscation problems arise in the context of privacy in cyber-physical and cloud-based systems \cite{Tanaka2017,Nekouei2019}, and covert navigation in robotics \cite{Marzouqi2011,Marzouqi2006}.

\begin{figure}
    \centering
    \includegraphics[width=0.9\columnwidth]{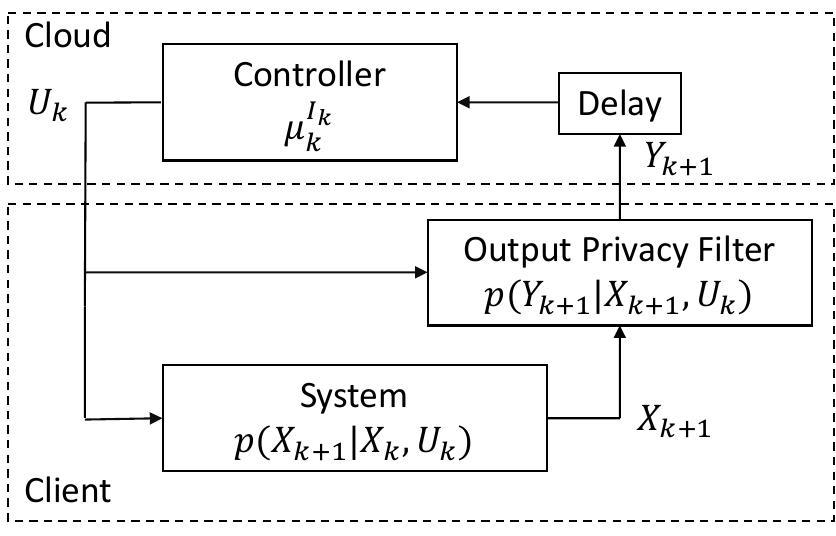}
    \caption{Cloud-based control scheme described in \cite{Tanaka2017,Nekouei2019}.}
    \label{fig:cloud_based}
\end{figure}

Consider again the setting shown in Fig. \ref{fig:activeEstimation}, but suppose now that the state trajectory estimator is owned by an adversary seeking to infer the state of the system.
The problem of controlling the system so as to hinder the adversary in inferring the state trajectory is consistent with \eqref{eq:activeEstimation} with $\beta < 0$ since increasing the smoother entropy corresponds to increasing the bounds \eqref{eq:smootherErrorBounds} on the adversary's ability to estimate the state trajectory using any estimator.
For example, consider the cloud-based control scheme illustrated in Fig.~\ref{fig:cloud_based} from \cite{Tanaka2017,Nekouei2019} in which a client seeks to have a system controlled by a cloud service without explicitly disclosing the system's state trajectory $X^T$.
The client provides the cloud service with outputs $Y_k$ of a privacy filter and the cloud service computes and returns control inputs $U_k$ using a policy provided by the client.
In the worst case (for the client), the cloud service also knows the system dynamics and the privacy filter.
The client is faced with the problem of controlling both the system and privacy filter to keep the state trajectory private whilst ensuring a suitable level of system performance, which is consistent with solving \eqref{eq:activeEstimation} with $\beta < 0$ (due to the relationship between the smoother entropy and estimation performance in \eqref{eq:smootherErrorBounds}).

\subsection{Comparison to Other State-Uncertainty Measures}

Our consideration of the smoother entropy for active estimation contrasts with approaches that instead minimise only the (marginal) entropy of the terminal state $H_\mu(X_T|Y^T, U^{T-1})$ (or equivalently, maximise the telescoping sum of information gains $\sum_{k = 0}^{T-1} [H_\mu(X_{k} | Y^{k}, U^{k-1}) - H_\mu(X_{k+1} | Y^{k+1}, U^k)]$) \cite{Thrun2005, Roy2005}.
It also contrasts with approaches that instead minimise the sum of (marginal) entropies of the states $H_\mu(X_k|Y^k, U^{k-1})$ \cite{Krishnamurthy2007, Krishnamurthy2016, Araya2010, Nardi2019} or $H_\mu(X_k|Y^T, U^{T-1})$ \cite{Stachniss2005, Valencia2012} for $0 \leq k \leq T$.
Specifically, approaches based on marginal entropies neglect correlations between consecutive states and so overestimate the trajectory uncertainty captured by the smoother entropy in the sense that
\begin{align}
\label{eq:entropyBounds}
\begin{split}
\sum_{k = 0}^T H_\mu(X_k | Y^{k}, U^{k-1})
&\geq \sum_{k = 0}^T H_\mu(X_k | Y^{T}, U^{T-1})\\
&\geq  H_\mu(X^T | Y^T, U^{T-1}),
\end{split}
\end{align}
with equality holding only when the states are (temporally) independent.
Indeed, the additive form of the smoother entropy we established in \eqref{eq:secondAdditiveForm} now shows that the terms neglected by marginal-entropy approaches are exactly the mutual informations between consecutive states, i.e.\ $I_\mu(X_k;X_{k-1}|Y^{k-1}, U^{k-1})$.
%This second form \eqref{eq:secondAdditiveForm} highlights that active estimation and obfuscation approaches based on optimising the sum of marginal state uncertainties (as described before \eqref{eq:entropyBounds}) fail to consider the information shared between consecutive states.
Minimisation of the smoother entropy in \eqref{eq:activeEstimation} via $\beta > 0$ explicitly exploits these temporal dependencies between consecutive states.

Furthermore, minimising the smoother entropy is, in general, not equivalent to maximising the conditional mutual information $I_\mu(X^T; Y^T | U^{T-1}) = H_\mu(X^T | U^{T-1}) - H_\mu(X^T | Y^T, U^{T-1})$, which is often the goal in controlled sensing and optimal Bayesian experimental design \cite{Hoffmann2010}.
For example, whilst maximising $I_\mu(X^T; Y^T | U^{T-1})$ increases the dependence between the states and measurements, the states themselves could become more uncertain due to the term $H_\mu(X^T | U^{T-1})$.
Indeed, the mutual information $I_\mu(X^T; Y^T | U^{T-1})$ is the \emph{reduction} in state uncertainty due to the measurements and controls (cf.\ \cite[p.\ 19]{Cover2006}) --- it is not an absolute measure of state uncertainty.

Finally, maximisation of the smoother entropy for active obfuscation contrasts with the approach proposed in \cite{Tanaka2017} of minimising the directed information $I_\mu(X^T \to Y^T \| U^{T-1})$.
Theorem \ref{theorem:directedInformation} shows that maximising the smoother entropy has the potential to yield superior obfuscation performance to only minimising the directed information since it both decreases the information gained from the observations, i.e.\ $I(X^T \to Y^T \| U^{T-1})$, \emph{and} increases the unpredictability of the state process, i.e.\ $H(X^T \| Y^{T-1}, U^{T-1})$.
In the context of obfuscation, our results in Lemma \ref{lemma:estConvex} and Theorem \ref{theorem:estConcave} showing that maximisation of the smoother entropy leads to concave cost and value functions in \eqref{eq:firstActiveEstimationMDP} for $\beta < 0$ are also surprising and significant because the maximisation of the sum of marginal entropies \eqref{eq:entropyBounds} and most other state-uncertainty measures leads to nonconcave cost and value functions \cite{Fehr2018}.

\section{Active State Trajectory Estimation and Obfuscation Simulations}
\label{sec:results}

We now simulate \eqref{eq:activeEstimation} for active estimation and obfuscation.

\subsection{Simulation Set-Up}
\begin{figure}[t!]
    %\captionsetup[subfigure]{aboveskip=-1.5pt,belowskip=0.5pt,labelformat=simple}
    %\renewcommand*{\thesubfigure}{\quad(\alph{subfigure})}
    \centering
    \begin{subfigure}{0.48\columnwidth}
         \centering
         \includegraphics[width=\textwidth]{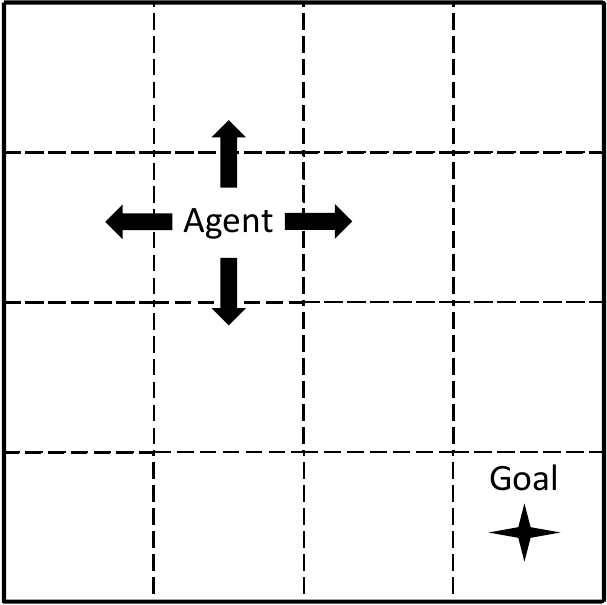}
         \caption{}
         \label{fig:agent}
     \end{subfigure}
     \hfill
     \begin{subfigure}{0.48\columnwidth}
         \centering
         \includegraphics[width=\textwidth]{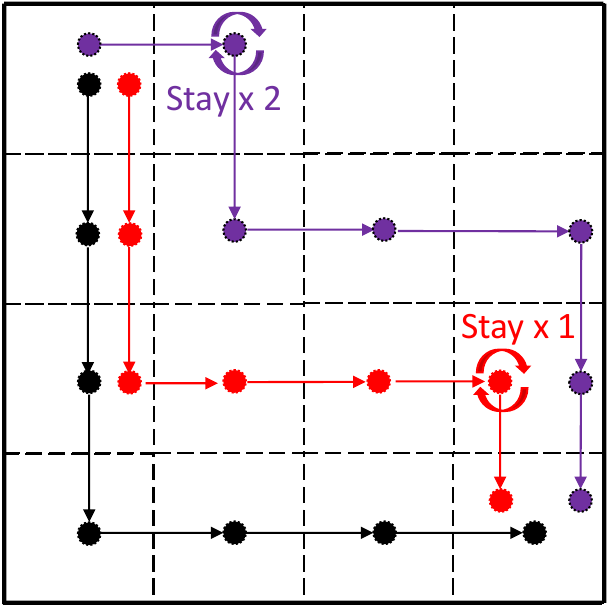}
         \caption{}
         \label{fig:activeEst}
     \end{subfigure}\\
     \begin{subfigure}{0.48\columnwidth}
         \centering
         \includegraphics[width=\textwidth]{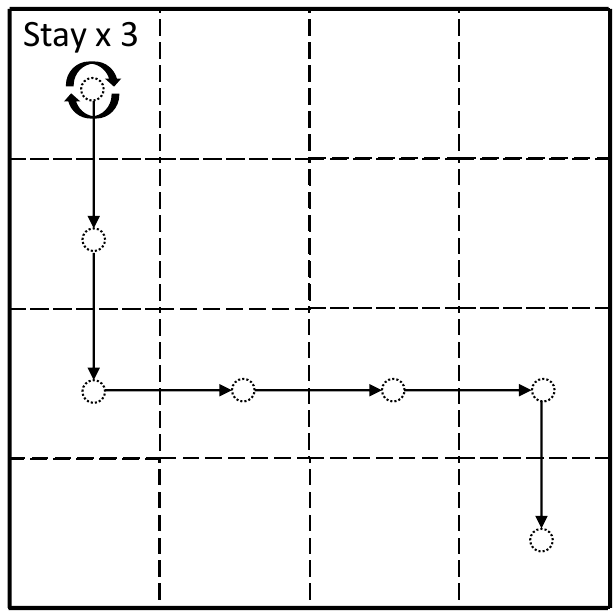}
         \caption{}
         \label{fig:activeEstComp}
     \end{subfigure}
     \hfill
     \begin{subfigure}{0.48\columnwidth}
         \centering
         \includegraphics[width=\textwidth]{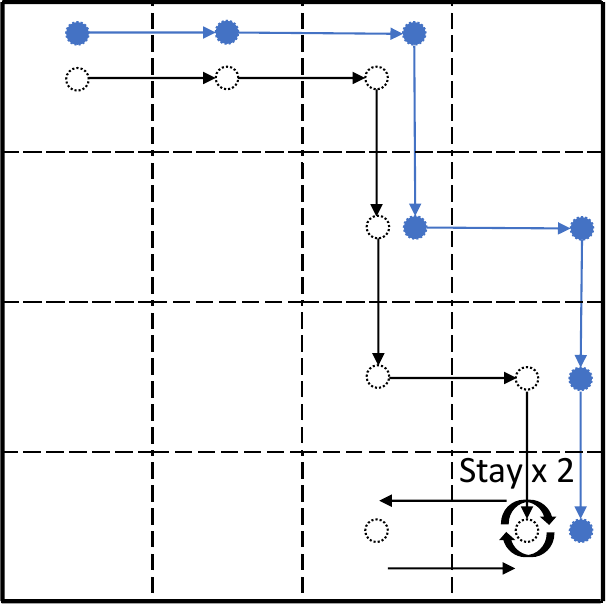}
         \caption{}
         \label{fig:activeObf}
     \end{subfigure}
    \caption{Simulation environment and example realisations. (a) Agent \& Goal. (b) Existing active estimation policies: Stand.\ POMDP policy (bottom in black); Min.\ Term.\ Ent.\ policy (middle in red); and, Min.\ Marg.\ Ent.\ policy (top in purple). (c) Our active estimation policy solving \eqref{eq:activeEstimation} with $\beta = 1$. (d) Active obfuscation policies: our active obfuscation policy solving \eqref{eq:activeEstimation} with $\beta = -1$ (bottom in black, Stay at $k = 6,7$); and, Min.\ Dir.\ Info. policy (top in blue). Transitions not shown correspond to remaining in goal cell.}
    \label{fig:navigation}
    %\vspace{-0.2cm}
\end{figure}

For the purpose of simulations, we consider a version of the benchmark \texttt{4x4.95} POMDP\footnote{https://www.pomdp.org/examples/} \cite{Parr1995,Littman1995} in which an agent moves in a $4 \times 4$ grid as shown in Fig.\ \ref{fig:agent}.
Each cell in the grid is taken as a state such that $\mathcal{X} = \{1, \ldots, 16\}$ (enumerated top-to-bottom, left-to-right).
There are five possible control actions $\mathcal{U} = \{1,2,3,4,5\}$ corresponding to the agent: transitioning to one of the four neighbouring cells left, right, up, or down with probability $0.8$ (failing to move with probability $0.2$); or, staying put with probability $1$. If a transition would take the agent out of the grid then it remains stationary.
We enlarge the measurement space compared to the benchmark \texttt{4x4.95} POMDP, with the agent given inexact information about its position in the grid through measurements $\mathcal{Y} = \{0,1,2,3,4\}$ corresponding to the number of walls detected adjacent to its current cell.
In each cell, the agent detects a wall when it is present with probability $0.9$, but detects a wall when it is not present with probability $0.1$.
The agent is initially placed (uniformly) randomly in one of the cells and is not provided with knowledge of this cell.
The horizon is $T = 10$.
Importantly, the state, measurement, and control space dimensions of this environment are similar to (or exceed) those of recent benchmark POMDPs for active estimation and active obfuscation problems in \cite{Fehr2018,Thomas2020}.

Inspired by uncertainty-aware robot navigation problems in which robots must plan and execute trajectories along which they are able to localise themselves (cf.\ \cite{Roy1999,Thrun2005,Nardi2019}), we first consider the agent to be seeking to reach the given goal in the bottom-right-most cell whilst ensuring that its path can be estimated for the purpose of later being retraced, communicated, or used for mapping.
We examine the ability of the agent to achieve these dual objectives by solving \eqref{eq:activeEstimation} with $\beta = 1$ and the costs $c_T(x_T) = \mathbbm{1}_{\{x_T \neq 16\}}$ and $c_k(x_k, u_k) = 0$ for all $x_k \in \mathcal{X}$ and $u_k \in \mathcal{U}$.
We specifically use the PWLC approximate solution approach detailed in Section \ref{sec:approx} with the approximation $\hat{g}_k^\beta$ computed using a set $\Xi$ with $\delta_\Xi = 0.92$ and containing the middle of the simplex $\Delta$ and points near the vertices with values in their largest element of $1 - 0.001(N_x-1)$ and $0.001$ in their other $N_x-1$ elements.
We use a standard POMDP solver\footnote{https://www.pomdp.org/code/} modified for PWLC costs (as detailed in \cite[Section 8.4.5]{Krishnamurthy2016}).
Using the same costs $c_T$ and $c_k$, we also simulate for comparison: a standard POMDP policy (\emph{Stand.\ POMDP}) solving \eqref{eq:activeEstimation} with $\beta = 0$; a minimum marginal entropy policy (\emph{Min.\ Marg.\ Ent.}) as in \cite{Krishnamurthy2007,Krishnamurthy2016,Araya2010} that minimises the sum of marginal entropies $\sum_{k = 0}^T H_\mu(X_k | Y^k, U^{k-1})$ instead of $H_\mu(X^T | Y^T, U^{T-1})$ in \eqref{eq:activeEstimation}; and,a minimum terminal entropy policy (\emph{Min.\ Term.\ Ent.}) as in \cite{Roy2005} that minimises $H_\mu(X_T | Y^T, U^{T-1})$ instead of $H_\mu(X^T | Y^T, U^{T-1})$ in \eqref{eq:activeEstimation}.

Inspired by covert robot navigation problems in which robots must plan and execute trajectories along which they are \emph{difficult} to track by potential adversaries \cite{Marzouqi2011,Hibbard2019}, we also simulate the agent as seeking to reach the given goal location whilst ensuring that its path is difficult to estimate.
To examine the ability of the agent to perform this active obfuscation, we solve \eqref{eq:activeEstimation} with $\beta = -1$ and the same costs $c_T$ and $c_k$ as before.
We again use the PWLC solution approach detailed in Section \ref{sec:approx} with the same base points and POMDP solver, but with the functions $\hat{\ell}_k^\beta$ (due to the concavity of $\ell_k^\beta$ for $\beta = -1$).
For comparison, we also simulate the minimum directed information policy (\emph{Min.\ Dir.\ Info.}) of \cite{Tanaka2017} that uses $I(X^T \to Y^T \| U^{T-1})$ instead of $-H(X^T | Y^T, U^{T-1})$ in \eqref{eq:activeEstimation}.

\subsection{Simulation Results}

Table \ref{tbl:navigation} summarises the terminal cost, smoother entropy, and probability of error of MAP state trajectory estimates (from the Viterbi algorithm) computed from $5000$ Monte Carlo simulations of each policy.
%The smoother entropies are estimated by averaging the entropies $H(X^T | y^{T}, u^{T-1})$ of the posterior state distributions $p(x^{T} | y^{T}, u^{T-1})$ over the Monte Carlo runs, whilst the MAP error probabilities are estimated by counting the number of times the Viterbi trajectory estimate differs from the true state trajectory (averaging over the Monte Carlo runs).
Example realisations are in Fig. \ref{fig:navigation}.

The results in Table \ref{tbl:navigation} suggest that the standard POMDP policy offers the lowest terminal cost since it moves the agent directly towards the goal without consideration of either active estimation or obfuscation (as illustrated in Fig.\ \ref{fig:activeEstComp}).
Our active estimation policy (\eqref{eq:activeEstimation} with $\beta = 1$) minimises the smoother entropy but has a greater terminal cost than the other policies.
Indeed, as illustrated in Fig.\ \ref{fig:activeEst}, our active estimation policy often reduces the uncertainty about the initial state $X_0$, and hence the entire trajectory, by initially electing to keep the agent still so as to receive measurements without changing the state.
Our active estimation policy elects only to move the agent after the initial state uncertainty is reduced, which leads to better trajectory estimates (as evidenced by the lesser MAP error probability in Table \ref{tbl:navigation}) but sometimes results in time being exhausted before the agent reaches the goal.
In contrast, the \emph{Min.\ Marg.\ Ent.} and \emph{Min.\ Term.\ Ent.} policies typically elect to move immediately and reduce instantaneous state uncertainties by keeping the agent still at isolated time instances $k > 0$ (see Fig.~\ref{fig:activeEstComp}).
The \emph{Min.\ Marg.\ Ent.} and \emph{Min.\ Term.\ Ent.} policies thus achieve lesser terminal costs but greater smoother entropies than our active estimation policy.

From Table \ref{tbl:navigation}, we also see that our active obfuscation policy (\eqref{eq:activeEstimation} with $\beta =  - 1$) increases the smoother entropy more than the \emph{Min.\ Dir.\ Info.} policy.
%As a consequence, the error probability of MAP estimates under our active obfuscation policy is greater than under the \emph{Min.\ Dir.\ Info.} policy.
The reason for this difference is that our active obfuscation policy increases both the unpredictability of the state process, i.e.\ $H_\mu(X^T \| Y^{T-1}, U^{T-1})$, and decreases the information gained from the measurements, i.e.\ $I_\mu(X^T \to Y^T \| U^{T-1})$ (cf.\ Theorem \ref{theorem:directedInformation} and Fig. \ref{fig:activeObf}).
In contrast, the \emph{Min.\ Dir.\ Info.} policy only decreases the information gained from the measurements.

Our simulations therefore suggest that optimising the smoother entropy via \eqref{eq:activeEstimation} with $\beta =  \pm 1$ can offer superior active estimation and obfuscation performance compared to optimising alternative state-uncertainty measures due to the smoother entropy capturing correlations between states.
Furthermore, the ability to both minimise and maximise the smoother entropy using standard POMDP techniques is unique amongst the other state-uncertainty measure considered in our simulations (the sum of marginal entropies, the terminal entropy, and the directed information) since no other leads to concave cost and value functions when both minimised and maximised.

\begin{table}[t!]
\begin{center}
\caption{Uncertainty-Aware and Covert Navigation: Terminal cost, Smoother entropy, and Maximum a posteriori (MAP) error probabilities. Minimum and Maximum (best for active estimation and obfuscation, resp.) values are in bold.}
\label{tbl:navigation}
\begin{tabular}{@{}lccc@{}}
\toprule
\multicolumn{1}{c}{\multirow{2}{*}{\textbf{Policy}}} & \textbf{Term. Cost} & \textbf{Smoother} & \multirow{2}{*}{\textbf{\begin{tabular}[c]{@{}c@{}}MAP Err.\\ Prob. \end{tabular}}} \\
\multicolumn{1}{c}{}            &                     $E[c_T(x_T)]$ & \textbf{Entropy}&                                                    \\ \cmidrule(rl){1-4} 
\textbf{Active Est.: \eqref{eq:activeEstimation} with $\beta = 1$}                & 0.182 & \textbf{1.701} & \textbf{0.495} \\
\textbf{Active Est.: Min. Marg. Ent.} & 0.164 & 1.779 & 0.516 \\
\textbf{Active Est.: Min. Term. Ent.} & 0.127 & 1.874 & 0.540 \\
\textbf{Stand. POMDP} \eqref{eq:activeEstimation} with $\beta = 0$ & \textbf{0.023} & 1.893 & 0.542 \\ 
\textbf{Active Obf.: Min. Dir. Info.} & 0.050 & 1.925 & 0.555 \\
\textbf{Active Obf.: \eqref{eq:activeEstimation} with $\beta = -1$} & 0.179 & \textbf{2.334} & \textbf{0.625} \\
\bottomrule
\end{tabular}
\end{center}
\end{table}

\section{Conclusion}
\label{sec:conclusion}

We investigated the smoother entropy (i.e.\ the conditional entropy of the state trajectory given measurements and controls) as a tractable criterion for active state estimation and obfuscation.
We established novel forms of the smoother entropy using the Marko-Massey theory of directed information that surprisingly enable both its minimisation and maximisation using standard POMDP techniques.
Future work could include investigating game-theoretic formulations of adversarial active estimation and obfuscation using the smoother entropy, including inverse problems such as inverse filtering (cf.\ \cite{Lourenco2020,Krishnamurthy2019,Mattila2020}).
%Similarly, as in recent work \cite{Lourenco2020}, it would be interesting to investigate optimising alternative privacy functions to the smoother entropy to conceal entire state trajectories.
%Future work will investigate reinforcement learning for solving our active estimation and obfuscation problems (which are undiscounted and finite-horizon whilst much reinforcement learning work is concerned with infinite-horizon discounted problems).
%\section*{Appendix}
%\label{sec:appendix}

% \section*{Acknowledgement}

%% References section
\bibliographystyle{IEEEtran}
%\balance
\bibliography{IEEEabrv,Library}

\begin{thebibliography}{10}
\providecommand{\url}[1]{#1}
\csname url@samestyle\endcsname
\providecommand{\newblock}{\relax}
\providecommand{\bibinfo}[2]{#2}
\providecommand{\BIBentrySTDinterwordspacing}{\spaceskip=0pt\relax}
\providecommand{\BIBentryALTinterwordstretchfactor}{4}
\providecommand{\BIBentryALTinterwordspacing}{\spaceskip=\fontdimen2\font plus
\BIBentryALTinterwordstretchfactor\fontdimen3\font minus
  \fontdimen4\font\relax}
\providecommand{\BIBforeignlanguage}[2]{{%
\expandafter\ifx\csname l@#1\endcsname\relax
\typeout{** WARNING: IEEEtran.bst: No hyphenation pattern has been}%
\typeout{** loaded for the language `#1'. Using the pattern for}%
\typeout{** the default language instead.}%
\else
\language=\csname l@#1\endcsname
\fi
#2}}
\providecommand{\BIBdecl}{\relax}
\BIBdecl

\bibitem{Molloy2021}
T.~L. Molloy and G.~N. Nair, ``Smoothing-averse control: Covertness and privacy
  from smoothers,'' in \emph{2021 American Control Conference (ACC)}, 2021, pp.
  4598--4605.

\bibitem{Molloy2021a}
------, ``Active trajectory estimation for partially observed {Markov} decision
  processes via conditional entropy,'' in \emph{2021 European Control
  Conference (ECC)}, 2021, pp. 385--391.

\bibitem{Blackmore2008}
L.~Blackmore, S.~Rajamanoharan, and B.~C. Williams, ``{Active estimation for
  jump Markov linear systems},'' \emph{IEEE Transactions on Automatic Control},
  vol.~53, no.~10, pp. 2223--2236, 2008.

\bibitem{Hu2004}
X.~Hu and T.~Ersson, ``Active state estimation of nonlinear systems,''
  \emph{Automatica}, vol.~40, no.~12, pp. 2075 -- 2082, 2004.

\bibitem{Baglietto2007}
M.~Baglietto, G.~Battistelli, and L.~Scardovi, ``Active mode observability of
  switching linear systems,'' \emph{Automatica}, vol.~43, no.~8, pp.
  1442--1449, 2007.

\bibitem{Scardovi2007}
L.~{Scardovi}, M.~{Baglietto}, and T.~{Parisini}, ``Active state estimation for
  nonlinear systems: A neural approximation approach,'' \emph{IEEE Transactions
  on Neural Networks}, vol.~18, no.~4, pp. 1172--1184, 2007.

\bibitem{Krishnamurthy2016}
V.~Krishnamurthy, \emph{Partially observed {Markov} decision processes}.\hskip
  1em plus 0.5em minus 0.4em\relax Cambridge University Press, 2016.

\bibitem{Zois2014}
D.-S. Zois, M.~Levorato, and U.~Mitra, ``{Active classification for POMDPs: A
  Kalman-like state estimator},'' \emph{IEEE Transactions on Signal
  Processing}, vol.~62, no.~23, pp. 6209--6224, 2014.

\bibitem{Zois2017}
D.-S. Zois and U.~Mitra, ``Active state tracking with sensing costs: Analysis
  of two-states and methods for $n$-states,'' \emph{IEEE Transactions on Signal
  Processing}, vol.~65, no.~11, pp. 2828--2843, 2017.

\bibitem{Chattopadhyay2018}
A.~Chattopadhyay and U.~Mitra, ``{Active sensing for Markov chain tracking},''
  in \emph{2018 IEEE Global Conference on Signal and Information Processing
  (GlobalSIP)}.\hskip 1em plus 0.5em minus 0.4em\relax IEEE, 2018, pp.
  1050--1054.

\bibitem{Krishnamurthy2020}
V.~{Krishnamurthy}, ``Convex stochastic dominance in {Bayesian} localization,
  filtering, and controlled sensing pomdps,'' \emph{IEEE Transactions on
  Information Theory}, vol.~66, no.~5, pp. 3187--3201, 2020.

\bibitem{Hoffmann2010}
G.~M. Hoffmann and C.~J. Tomlin, ``Mobile sensor network control using mutual
  information methods and particle filters,'' \emph{IEEE Transactions on
  Automatic Control}, vol.~55, no.~1, pp. 32--47, 2010.

\bibitem{Kartik2018}
D.~Kartik, E.~Sabir, U.~Mitra, and P.~Natarajan, ``Policy design for active
  sequential hypothesis testing using deep learning,'' in \emph{2018 56th
  Annual Allerton Conference on Communication, Control, and Computing
  (Allerton)}.\hskip 1em plus 0.5em minus 0.4em\relax IEEE, 2018, pp. 741--748.

\bibitem{Mu2016}
B.~Mu, M.~Giamou, L.~Paull, A.-a. {Agha-Mohammadi}, J.~Leonard, and J.~How,
  ``Information-based active {SLAM} via topological feature graphs,'' in
  \emph{2016 IEEE 55th Conference on Decision and Control (CDC)}.\hskip 1em
  plus 0.5em minus 0.4em\relax IEEE, 2016, pp. 5583--5590.

\bibitem{Thrun2005}
S.~Thrun, W.~Burgard, and D.~Fox, \emph{Probabilistic Robotics}.\hskip 1em plus
  0.5em minus 0.4em\relax MIT Press, 2005.

\bibitem{Roy1999a}
N.~Roy, W.~Burgard, D.~Fox, and S.~Thrun, ``Coastal navigation-mobile robot
  navigation with uncertainty in dynamic environments,'' in \emph{Proceedings
  1999 IEEE International Conference on Robotics and Automation (ICRA)}.\hskip
  1em plus 0.5em minus 0.4em\relax IEEE, 1999, pp. 35--40.

\bibitem{Stachniss2005}
C.~Stachniss, G.~Grisetti, and W.~Burgard, ``Information gain-based exploration
  using {Rao-Blackwellized} particle filters.'' in \emph{Robotics: Science and
  Systems}, vol.~2, 2005, pp. 65--72.

\bibitem{Valencia2012}
R.~Valencia, J.~Valls~Miró, G.~Dissanayake, and J.~Andrade-Cetto, ``{Active
  Pose SLAM},'' in \emph{2012 IEEE/RSJ International Conference on Intelligent
  Robots and Systems}, 2012, pp. 1885--1891.

\bibitem{Roy2005}
R.~{Sim} and N.~{Roy}, ``{Global A-Optimal Robot Exploration in SLAM},'' in
  \emph{Proceedings of the 2005 IEEE International Conference on Robotics and
  Automation}, 2005, pp. 661--666.

\bibitem{Li2018}
S.~{Li}, A.~{Khisti}, and A.~{Mahajan}, ``Information-theoretic privacy for
  smart metering systems with a rechargeable battery,'' \emph{IEEE Trans. on
  Information Theory}, vol.~64, no.~5, pp. 3679--3695, 2018.

\bibitem{Li2019}
N.~Li, I.~Kolmanovsky, and A.~Girard, ``Detection-averse optimal and
  receding-horizon control for {Markov} decision processes,''
  \emph{Automatica}, vol. 122, p. 109278, 2020.

\bibitem{Farokhi2020}
F.~Farokhi, Ed., \emph{Privacy in Dynamical Systems}.\hskip 1em plus 0.5em
  minus 0.4em\relax Springer, 2020.

\bibitem{Tanaka2017}
T.~{Tanaka}, M.~{Skoglund}, H.~{Sandberg}, and K.~H. {Johansson}, ``Directed
  information and privacy loss in cloud-based control,'' in \emph{2017 American
  Control Conference (ACC)}, 2017, pp. 1666--1672.

\bibitem{Savas2020}
Y.~{Savas}, M.~{Ornik}, M.~{Cubuktepe}, M.~O. {Karabag}, and U.~{Topcu},
  ``Entropy maximization for {Markov} decision processes under temporal logic
  constraints,'' \emph{IEEE Trans. on Automatic Control}, vol.~65, no.~4, pp.
  1552--1567, 2020.

\bibitem{Shateri2020}
M.~{Shateri}, F.~{Messina}, P.~{Piantanida}, and F.~{Labeau}, ``Real-time
  privacy-preserving data release for smart meters,'' \emph{IEEE Transactions
  on Smart Grid}, vol.~11, no.~6, pp. 5174--5183, 2020.

\bibitem{Marzouqi2011}
M.~S. Marzouqi and R.~A. Jarvis, ``Robotic covert path planning: A survey,'' in
  \emph{2011 IEEE 5th international conference on robotics, automation and
  mechatronics (RAM)}.\hskip 1em plus 0.5em minus 0.4em\relax IEEE, 2011, pp.
  77--82.

\bibitem{Hibbard2019}
M.~{Hibbard}, Y.~{Savas}, B.~{Wu}, T.~{Tanaka}, and U.~{Topcu}, ``Unpredictable
  planning under partial observability,'' in \emph{2019 IEEE 58th Conference on
  Decision and Control (CDC)}, 2019, pp. 2271--2277.

\bibitem{Haugh2020}
M.~B. Haugh and O.~R. Lacedelli, ``{Information Relaxation Bounds for Partially
  Observed Markov Decision Processes},'' \emph{IEEE Transactions on Automatic
  Control}, vol.~65, no.~8, pp. 3256--3271, 2020.

\bibitem{Walraven2019}
E.~Walraven and M.~T. Spaan, ``Point-based value iteration for finite-horizon
  {POMDPs},'' \emph{Journal of Artificial Intelligence Research}, vol.~65, pp.
  307--341, 2019.

\bibitem{Kurniawati2008}
H.~Kurniawati, D.~Hsu, and W.~S. Lee, ``{SARSOP}: Efficient point-based {POMDP}
  planning by approximating optimally reachable belief spaces.'' in
  \emph{Robotics: Science and systems}, vol. 2008.\hskip 1em plus 0.5em minus
  0.4em\relax Zurich, Switzerland., 2008.

\bibitem{Garg2019}
N.~P. Garg, D.~Hsu, and W.~S. Lee, ``{DESPOT-Alpha}: Online {POMDP} planning
  with large state and observation spaces.'' in \emph{Robotics: Science and
  Systems}, 2019.

\bibitem{Krishnamurthy2007}
V.~Krishnamurthy and D.~V. Djonin, ``Structured threshold policies for dynamic
  sensor scheduling -- a partially observed {Markov} decision process
  approach,'' \emph{IEEE Transactions on Signal Processing}, vol.~55, no.~10,
  pp. 4938--4957, 2007.

\bibitem{Araya2010}
M.~Araya, O.~Buffet, V.~Thomas, and F.~Charpillet, ``A {POMDP} extension with
  belief-dependent rewards,'' in \emph{Advances in Neural Information
  Processing Systems}, J.~Lafferty, C.~Williams, J.~Shawe-Taylor, R.~Zemel, and
  A.~Culotta, Eds., vol.~23.\hskip 1em plus 0.5em minus 0.4em\relax Curran
  Associates, Inc., 2010, pp. 64--72.

\bibitem{Flayac2017}
E.~{Flayac}, K.~{Dahia}, B.~{Hérissé}, and F.~{Jean}, ``Nonlinear {Fisher}
  particle output feedback control and its application to terrain aided
  navigation,'' in \emph{2017 IEEE 56th Annual Conference on Decision and
  Control (CDC)}, 2017, pp. 1566--1571.

\bibitem{Bar-Shalom2001}
Y.~Bar-Shalom, X.~Rong~Li, and T.~Kirubarajan, \emph{{Estimation with
  applications to tracking and navigation}}.\hskip 1em plus 0.5em minus
  0.4em\relax New York, NY: John Wiley \& Sons, 2001.

\bibitem{Sandberg2015}
H.~{Sandberg}, G.~{Dán}, and R.~{Thobaben}, ``Differentially private state
  estimation in distribution networks with smart meters,'' in \emph{54th IEEE
  Conference on Decision and Control (CDC)}, 2015, pp. 4492--4498.

\bibitem{Hale2015}
M.~Hale and M.~Egerstedt, ``Differentially private cloud-based multi-agent
  optimization with constraints,'' in \emph{2015 American Control Conference
  (ACC)}.\hskip 1em plus 0.5em minus 0.4em\relax IEEE, 2015, pp. 1235--1240.

\bibitem{Nekouei2019}
E.~Nekouei, T.~Tanaka, M.~Skoglund, and K.~H. Johansson,
  ``Information-theoretic approaches to privacy in estimation and control,''
  \emph{Annual Reviews in Control}, 2019.

\bibitem{Murguia2021}
C.~{Murguia}, I.~{Shames}, F.~{Farokhi}, D.~{Nešić}, and H.~V. {Poor}, ``On
  privacy of dynamical systems: An optimal probabilistic mapping approach,''
  \emph{IEEE Transactions on Information Forensics and Security}, vol.~16, pp.
  2608--2620, 2021.

\bibitem{Fehr2018}
M.~Fehr, O.~Buffet, V.~Thomas, and J.~Dibangoye, ``{$\rho$-POMDPs have
  Lipschitz-Continuous epsilon-Optimal Value Functions},'' in \emph{Advances in
  Neural Information Processing Systems}, S.~Bengio, H.~Wallach, H.~Larochelle,
  K.~Grauman, N.~Cesa-Bianchi, and R.~Garnett, Eds., vol.~31.\hskip 1em plus
  0.5em minus 0.4em\relax Curran Associates, Inc., 2018.

\bibitem{Feder1994}
M.~{Feder} and N.~{Merhav}, ``Relations between entropy and error
  probability,'' \emph{IEEE Transactions on Information Theory}, vol.~40,
  no.~1, pp. 259--266, 1994.

\bibitem{Marko1973}
H.~{Marko}, ``The bidirectional communication theory - a generalization of
  information theory,'' \emph{IEEE Trans. on Communications}, vol.~21, no.~12,
  pp. 1345--1351, 1973.

\bibitem{Massey1990}
J.~Massey, ``Causality, feedback and directed information,'' in \emph{Proc.
  Int. Symp. Inf. Theory Applic.(ISITA-90)}, 1990, pp. 303--305.

\bibitem{Kramer1998}
G.~Kramer, \emph{Directed information for channels with feedback}.\hskip 1em
  plus 0.5em minus 0.4em\relax Hartung-Gorre, 1998.

\bibitem{Massey2005}
J.~L. {Massey} and P.~C. {Massey}, ``Conservation of mutual and directed
  information,'' in \emph{Proceedings. International Symposium on Information
  Theory, 2005. ISIT 2005.}, 2005, pp. 157--158.

\bibitem{Cover2006}
T.~Cover and J.~Thomas, \emph{{Elements of information theory}}, 2nd~ed.\hskip
  1em plus 0.5em minus 0.4em\relax New York: Wiley, 2006.

\bibitem{Bar1974}
Y.~Bar-Shalom and E.~Tse, ``Dual effect, certainty equivalence, and separation
  in stochastic control,'' \emph{IEEE Transactions on Automatic Control},
  vol.~19, no.~5, pp. 494--500, 1974.

\bibitem{Briers2010}
M.~Briers, A.~Doucet, and S.~Maskell, ``Smoothing algorithms for state--space
  models,'' \emph{Annals of the Institute of Statistical Mathematics}, vol.~62,
  no.~1, p.~61, 2010.

\bibitem{Lee2021}
Y.~Lee, P.~Cai, and D.~Hsu, ``{MAGIC: Learning Macro-Actions for Online POMDP
  Planning },'' in \emph{Proceedings of Robotics: Science and Systems},
  Virtual, July 2021.

\bibitem{Haarnoja2017}
T.~Haarnoja, H.~Tang, P.~Abbeel, and S.~Levine, ``Reinforcement learning with
  deep energy-based policies,'' in \emph{{International Conference on Machine
  Learning}}.\hskip 1em plus 0.5em minus 0.4em\relax PMLR, 2017, pp.
  1352--1361.

\bibitem{Haarnoja2018}
T.~Haarnoja, A.~Zhou, P.~Abbeel, and S.~Levine, ``Soft actor-critic: Off-policy
  maximum entropy deep reinforcement learning with a stochastic actor,'' in
  \emph{{International Conference on Machine Learning}}.\hskip 1em plus 0.5em
  minus 0.4em\relax PMLR, 2018, pp. 1861--1870.

\bibitem{Valencia2018}
R.~Valencia and J.~Andrade-Cetto, ``{Active Pose SLAM},'' in \emph{Mapping,
  Planning and Exploration with Pose SLAM}.\hskip 1em plus 0.5em minus
  0.4em\relax Springer, 2018, pp. 89--108.

\bibitem{Hernando2005}
D.~{Hernando}, V.~{Crespi}, and G.~{Cybenko}, ``Efficient computation of the
  hidden {Markov} model entropy for a given observation sequence,'' \emph{IEEE
  Transactions on Information Theory}, vol.~51, no.~7, pp. 2681--2685, 2005.

\bibitem{Bertsekas2005}
D.~P. Bertsekas, \emph{Dynamic programming and optimal control},
  {Third}~ed.\hskip 1em plus 0.5em minus 0.4em\relax Belmont, MA: Athena
  Scientific, 1995, vol.~1.

\bibitem{Globerson2007}
A.~Globerson and T.~Jaakkola, ``Approximate inference using conditional entropy
  decompositions,'' in \emph{Artificial Intelligence and Statistics}, 2007, pp.
  131--138.

\bibitem{Downarowicz2011}
T.~Downarowicz, \emph{Entropy in dynamical systems}.\hskip 1em plus 0.5em minus
  0.4em\relax Cambridge University Press, 2011, vol.~18.

\bibitem{Fiorenza2017}
R.~Fiorenza, \emph{H{\"o}lder and locally H{\"o}lder Continuous Functions, and
  Open Sets of Class $C^k$, $C^{k, \lambda}$}.\hskip 1em plus 0.5em minus
  0.4em\relax Birkh{\"a}user, 2017.

\bibitem{Bertsekas2012}
D.~P. Bertsekas, \emph{Dynamic programming and optimal control},
  {Fourth}~ed.\hskip 1em plus 0.5em minus 0.4em\relax Belmont, MA: Athena
  Scientific, 2012, vol.~2.

\bibitem{Hauskrecht2000}
M.~Hauskrecht, ``Value-function approximations for partially observable
  {Markov} decision processes,'' \emph{Journal of Artificial Intelligence
  Research}, vol.~13, pp. 33--94, 2000.

\bibitem{Hoerger2021}
M.~Hoerger and H.~Kurniawati, ``{An On-Line POMDP Solver for Continuous
  Observation Spaces},'' in \emph{2021 IEEE International Conference on
  Robotics and Automation (ICRA)}, 2021, pp. 7643--7649.

\bibitem{Nardi2019}
L.~Nardi and C.~Stachniss, ``{Uncertainty-Aware Path Planning for Navigation on
  Road Networks Using Augmented MDPs},'' in \emph{2019 International Conference
  on Robotics and Automation (ICRA)}.\hskip 1em plus 0.5em minus 0.4em\relax
  IEEE, 2019, pp. 5780--5786.

\bibitem{Elliott1995}
R.~Elliott, L.~Aggoun, and J.~Moore, \emph{{Hidden Markov Models: Estimation
  and Control}}.\hskip 1em plus 0.5em minus 0.4em\relax New York, NY: Springer,
  1995.

\bibitem{Kaess2008}
M.~Kaess, A.~Ranganathan, and F.~Dellaert, ``{iSAM}: Incremental smoothing and
  mapping,'' \emph{IEEE Transactions on Robotics}, vol.~24, no.~6, pp.
  1365--1378, 2008.

\bibitem{Marzouqi2006}
M.~S. Marzouqi and R.~A. Jarvis, ``New visibility-based path-planning approach
  for covert robotic navigation,'' \emph{Robotica}, vol.~24, no.~6, pp.
  759--773, 2006.

\bibitem{Parr1995}
R.~Parr and S.~Russell, ``Approximating optimal policies for partially
  observable stochastic domains,'' in \emph{Proceedings of the 14th
  International Joint Conference on Artificial Intelligence - Volume 2}, ser.
  IJCAI'95.\hskip 1em plus 0.5em minus 0.4em\relax San Francisco, CA, USA:
  Morgan Kaufmann Publishers Inc., 1995, p. 1088–1094.

\bibitem{Littman1995}
M.~L. Littman, A.~R. Cassandra, and L.~P. Kaelbling, ``Learning policies for
  partially observable environments: Scaling up,'' in \emph{Machine Learning
  Proceedings 1995}.\hskip 1em plus 0.5em minus 0.4em\relax Elsevier, 1995, pp.
  362--370.

\bibitem{Thomas2020}
V.~Thomas, G.~Hutin, and O.~Buffet, ``Monte carlo information-oriented
  planning,'' in \emph{24th ECAI 2020-European Conference on Artificial
  Intelligence}, 2020.

\bibitem{Roy1999}
N.~Roy and S.~Thrun, ``Coastal navigation with mobile robots,'' in
  \emph{Proceedings of the 12th International Conference on Neural Information
  Processing Systems}, 1999, pp. 1043--1049.

\bibitem{Lourenco2020}
I.~{Lourenço}, R.~{Mattila}, C.~R. {Rojas}, and B.~{Wahlberg}, ``How to
  protect your privacy? a framework for counter-adversarial decision making,''
  in \emph{59th IEEE Conference on Decision and Control (CDC)}, 2020, pp.
  1785--1791.

\bibitem{Krishnamurthy2019}
V.~{Krishnamurthy} and M.~{Rangaswamy}, ``{How to Calibrate Your Adversary's
  Capabilities? Inverse Filtering for Counter-Autonomous Systems},'' \emph{IEEE
  Transactions on Signal Processing}, vol.~67, no.~24, pp. 6511--6525, 2019.

\bibitem{Mattila2020}
R.~{Mattila}, C.~R. {Rojas}, V.~{Krishnamurthy}, and B.~{Wahlberg}, ``{Inverse
  Filtering for Hidden Markov Models With Applications to Counter-Adversarial
  Autonomous Systems},'' \emph{IEEE Transactions on Signal Processing},
  vol.~68, pp. 4987--5002, 2020.

\end{thebibliography}

% \begin{IEEEbiography}
%     []{Timothy L. Molloy}
% \end{IEEEbiography}

% \begin{IEEEbiography}
%     []{Girish N. Nair}
% \end{IEEEbiography}

\end{document}